\documentclass[10pt,letterpaper]{article}
\usepackage[top=0.85in,left=2.75in,footskip=0.75in]{geometry}

\usepackage{amsmath,amssymb}

\usepackage{changepage}

\usepackage[utf8x]{inputenc}

\usepackage{textcomp,marvosym}

\usepackage{cite}

\usepackage{nameref,hyperref}

\usepackage[right]{lineno}

\usepackage{microtype}
\DisableLigatures[f]{encoding = *, family = * }

\usepackage[table]{xcolor}

\usepackage{array}

\usepackage{amsfonts,amsopn}
\usepackage{graphicx}
\usepackage{xspace}
\usepackage{subfig} 
\usepackage{mathtools} 
\usepackage{url}

\newcolumntype{+}{!{\vrule width 2pt}}

\newlength\savedwidth

\raggedright
\usepackage[aboveskip=1pt,labelfont=bf,labelsep=period,justification=raggedright,singlelinecheck=off]{caption}

\makeatletter
\renewcommand{\@biblabel}[1]{\quad#1.}
\makeatother

\usepackage{lastpage,fancyhdr,graphicx}
\usepackage{epstopdf}
\fancyheadoffset[L]{2.25in}
\fancyfootoffset[L]{2.25in}
\newcommand{\dfn}{\triangleq}
\newcommand{\untsph}{\mathbb{S}^{2}} 
\newcommand{\shc}[3]{{#1}_{#2}^{#3}}
\newcommand{\lsph}{L^2(\untsph)}
\newcommand{\conj}[1]{\overline{#1}} 

\newcommand{\intsph}{\int_{\untsph}}
\newcommand{\intphi}{\int_{0}^{2\pi}}
\newcommand{\figref}[1]{Fig \, \ref{#1}}
\newcommand{\nps}{\ensuremath{N_\textrm{O}}}
\newcommand{\bv}[1]{\boldsymbol{#1}}
\newcommand{\plms}{\widetilde{P}}

\newcommand{\basisGen}{{\cal K}}
\newcommand{\coeffGenVec}{\mathbf{x}}
\newcommand{\coeffSHVec}{\mathbf{c}}
\newcommand{\coeffSH}{c^m_\ell}
\newcommand{\coeffSPFVec}{\mathbf{e}}
\newcommand{\coeffSPF}{e_{n\ell m}}
\newcommand{\diffSigVector}{\mathbf{d}}
\newcommand{\SHorderVecPerShellAndn}{(\mathbf{c}_m)_{ns}}

\newtheorem{remark}{Remark}

\graphicspath{{figs/},{pdfs/},{pdf/},{Figures/},{tikz-figs/}}

\begin{document}
\vspace*{0.2in}

\begin{flushleft}
{\Large
\textbf\newline{Efficient sampling and robust 3D diffusion magnetic resonance imaging signal reconstruction} 
}
\newline
\\
Alice P. Bates\textsuperscript{1,2*},
Zubair Khalid\textsuperscript{3},
Jason D. McEwen\textsuperscript{4},
Rodney A. Kennedy\textsuperscript{1},
Alessandro Daducci\textsuperscript{5\Yinyang },
Erick J. Canales-Rodr\'{i}guez\textsuperscript{2,6,7,8\Yinyang},
\\
\bigskip
\textbf{1} Research School of Engineering, The Australian National University, Canberra, Australia
\\
\textbf{2} Signal Processing Lab (LTS5), \'{E}cole Polytechnique F\'{e}d\'{e}rale de Lausanne, Lausanne, Switzerland
\\
\textbf{3} Department of Electrical Engineering, School of Science and Engineering, Lahore University of Management Sciences, Lahore 54792, Pakistan
\\
\textbf{4} Mullard Space Science Laboratory, University College London, Surrey RH5 6NT, UK
\\
\textbf{5} Computer Science department, University of Verona, Verona, Italy
\\
\textbf{6} Department of Radiology, Centre Hospitalier Universitaire Vaudois, Lausanne, Switzerland
\\
\textbf{7} FIDMAG Germanes Hospital\'{a}ries, Sant Boi de Llobregat, Barcelona, Spain
\\
\textbf{8} Instituto de Salud Carlos III, Centro de Investigaci\'{o}n Biom\'{e}dica en Red de Salud Mental (CIBERSAM), Madrid, Spain
\\
\bigskip

%
%
\Yinyang These authors contributed equally to this work.





* alice.bates@anu.edu.au

\end{flushleft}
\section*{Abstract}
This paper presents novel single and multi-shell sampling schemes for diffusion MRI.  In diffusion MRI, it is paramount that the number of samples is as small as possible in order that scan times are practical in a clinical setting. The proposed schemes use an efficient number of measurements in that the number of samples is equal to the degrees of freedom in the orthonormal bases used for reconstruction. Novel reconstruction algorithms based on smaller subsystems of linear equations, as compared to the standard regularized least-squares method, are developed for both single and multi-shells sampling schemes. The smaller matrices used in these novel reconstruction algorithms are designed to be well-conditioned, leading to improved reconstruction accuracy. Accurate and robust reconstruction is also achieved through incorporation of regularization into the novel reconstruction algorithms and using a Rician or non-central Chi noise model. We quantitatively validate our single and multi-shell schemes against standard least-squares reconstruction methods to show that they enable more accurate reconstruction when the number of samples is equal to the degrees of freedom in the basis. Human brain data is also used to qualitatively evaluate reconstruction.

\section*{Introduction} 

Diffusion MRI is the preferred imaging modality for studying white-matter connectivity in the brain and diagnosing white-matter disorders~\cite{Assaf:2017,Bihan:2015}. Measurements of the diffusion signal are normally collected on single or multiple shells in $\mathbf{q}$-space.  It is necessary that the number of samples is as small as possible in order that scan times are suitable for a clinical setting, where acquisition time and cost need to be minimized~\cite{Assemlal:2011,Caruyer:2012b}.

From the $\mathbf{q}$-space samples, the intra-voxel diffusion signal can be analyzed using parametric approaches based on statistical~\cite{Basser:1994} and biophysical models~\cite{Jelescu:2017} as well as model-free reconstructions by expansion in an orthonormal basis~\cite{Novikov:2018,Daducci:2014}. One advantage of the latter approach is that it requires fewer assumptions on the nature of the diffusion signal. The only assumption being that the signal is band-limited in the corresponding basis: there is a maximum order on that basis above which there is little energy. This is unlike sparse representations of the diffusion signal~\cite{Ning:2015}, which have additional assumptions on the signal properties.  A number of local microstructure properties of the white matter tissue can be estimated from the reconstructed signal, including the diffusion orientation distribution function (ODF)~\cite{Aganj:2010,Cheng:2010}, fiber ODF~\cite{Canales:2015} and generalized fractional anisotropy~\cite{Cheng:2010,Tuch:2002}. Fiber tractography algorithms are employed to reconstruct the neural tracks connecting different brain regions from the local ODFs.

Accurate reconstruction of the diffusion signal in an orthonormal basis, with the only assumption being that a signal is band-limited in that basis, requires that the sampling grid is designed such that the number of samples is at least equal to the number of coefficients in the orthonormal basis and that their structure enables an accurate reconstruction transform to be defined~\cite{Khalid:2014,McEwen2011}. For accurate reconstruction of the diffusion signal, the effect of noise on the signal also needs to be modeled and its contribution removed from the measurements. Although the noise associated with the signal in each coil is Gaussian, the distribution of the diffusion signal depends on factors such as how the signals from these coils are combined and the number of coils. In order to remove phase related artifacts, often the magnitude of the diffusion MRI signal is taken. The Rician distribution arises when the magnitude of the complex signal from one coil or the magnitude of the sum of the signals from multiple coils is taken, while the non-central Chi distributions~(NCC) results from taking the root sum-of-squares of the complex signal from the different coils~\cite{Gudbjartsson:1995,Canales:2015,Varadarajan:2015}. NCC and Rician distributions are well-approximated as Gaussian if the signal-to-noise (SNR) of the data is high; when this is not the case, treating the noise as Gaussian degrades performance~\cite{Canales:2015,Varadarajan:2015}.

As discussed above, a sampling scheme and associated reconstruction algorithm should meet the following requirements: (1) as few measurements as possible to minimize scan time; (2) sampling grids are designed to enable accurate reconstruction; and (3) an accurate and robust reconstruction algorithm which takes into account the non-Gaussian distribution of the noise.

Most diffusion MRI sampling schemes use uniform sampling within each shell to ensure that the reconstruction accuracy is rotationally invariant~\cite{Caruyer:2011,Caruyer:2013,Cheng:2018,Leistedt:2012,Ye:2012}. Such schemes have a structure that only allows for least-squares reconstruction which requires more samples than the number of coefficients, as the least-squares matrix is ill-conditioned with this number of samples, and the accuracy improves with more measurements~\cite{Khalid:2014}. The standard reconstruction algorithm is regularized least-squares, which implicitly assumes Gaussian distributed noise. Regularization is used so that the reconstructed signal does not fit too closely to noisy measurements by trading off a smooth solution and fitting to the data~\cite{Descoteaux:2006,Caruyer:2012}. Another benefit of regularization is to improve the condition number of the matrix, which allows inverting the system with more stability.

Sampling schemes that use least-squares reconstruction can be used in the majorize-minimize framework proposed in~\cite{Varadarajan:2015}. The majorize-minimize framework iteratively finds the penalized maximum likelihood~(PML) estimate for Rician and NCC noise. The algorithm is a two step process where there is a measurement update to remove the noise bias and a coefficient update using regularized least-square to the noise variance. The Robust and Unbiased Model-BAsed spherical Deconvolution (RUMBA-SD) technique~\cite{Canales:2015}, that is based on a Richardson-Lucy algorithm adapted for Rician and NCC likelihood models, has a similar iterative algorithm for finding the maximum likelihood estimate for NCC noise for recovering the fiber ODF.

The single-shell sampling scheme proposed in~\cite{Bates:2016,Bates:2015c} has a number of samples equal to the number of coefficients in the orthonormal basis. The sampling grid in this scheme is designed so that it enables an accurate reconstruction algorithm with better conditioned matrices than the least-squares matrix. For this reason, this scheme enables more accurate reconstruction than the state-of-the-art sampling schemes which use least-squares for synthetic noise-free measurements of the diffusion signal~\cite{Bates:2015c}. Although the sampling scheme is not uniform, it does not have dense sampling on any region of the sphere and was found to achieve rotationally invariant reconstruction accuracy~\cite{Bates:2016,Bates:2015c}. However the scheme proposed in~\cite{Bates:2016} does not regularize the reconstruction and implicitly assumes that the noise has a Gaussian distribution.

This work extends~\cite{Bates:2016} to a multi-shell sampling scheme and aims to enable robust reconstruction of the diffusion signal by including regularization in the reconstruction algorithm and considering the non-Gaussian noise of the data, thereby meeting all three requirements. Preliminary parts of this work were reported in a conference paper~\cite{Bates:2017}.

This paper is organized as follows. First the notation and mathematical background is established before we present our proposed sampling and reconstruction schemes for single and multi-shell acquisitions respectively. We first present the single and multi-shell sampling grids and reconstruction algorithms before presenting the regularized and non-Gaussian noise removal extensions to these reconstruction algorithms. We then quantitatively evaluate our proposed schemes using synthetic data-sets and also perform qualitative evaluation using human brain data. 


\section*{Notation and mathematical background} \label{Sec:Math_Prelim}
Here we present the notation and mathematical formulation for the diffusion signal on a single shell (on the unit sphere) and multiple shells (in 3-dimensional space) in $\mathbf{q}$-space.


\subsection*{Diffusion signal on the sphere and spherical harmonics}
Let the diffusion signal at a fixed $\mathbf q$-space radius be denoted by $d(\theta,\phi)$, where the two angles colatitude $\theta \in [0, \pi]$ and longitude $\phi \in [0, 2\pi)$ parametrize a point $\bv{q}(\theta, \phi) = \bv{q}/||\bv{q}||_2 = (\sin\theta\cos\phi,\, \sin\theta\sin\phi,\, \cos \theta)' \in \mathbb{R}^3$ on the unit sphere $\untsph$.

The spherical harmonic (SH) functions form a complete basis for square integrable functions on the unit sphere $\lsph$, they are defined as~\cite{Kennedy-book:2013}
\begin{equation}
\label{Eq:Sph_harmonics} Y_{\ell}^{m}(\theta,
\phi) =
\sqrt{\frac{2{\ell}+1}{4\pi}\frac{({\ell}-m)!}{({\ell}+m)!}}\,
        P_{\ell}^{m}(\cos\theta)e^{im\phi},
\end{equation}
for integer degree ${\ell} \geq 0$ and integer order $ |m| \leq {\ell}$. In Eq~\eqref{Eq:Sph_harmonics}, $P_{\ell}^{m}$ denotes the associated Legendre function of degree $\ell$ and order $m$~\cite{Kennedy-book:2013}.
The SH coefficients of the diffusion signal on the sphere  $\coeffSH$ are given by the SH transform (SHt), defined as
\begin{equation}\label{Eq:dcoeff}
\coeffSH \dfn \intsph d(\theta,\phi) \conj {Y_{\ell}^m(\theta,\phi)}
\,\sin\theta\, d\theta\, d\phi.
\end{equation}
We can then reconstruct the diffusion signal on the sphere from its SH coefficients using the inverse SHt, given by
\begin{equation}
\label{Eq:d_expansion}
    d(\theta,\phi)=\sum_{\ell=0}^{\infty}\sum_{m=-{\ell}}^{\ell} \coeffSH  Y_{\ell}^m(\theta,\phi).
\end{equation}

The diffusion signal has the property that it is antipodally symmetric; it has the same value at locations diametrically opposite each other, with $d(\theta,\phi) = d(\pi - \theta,\phi + \pi)$. Since $Y_{\ell}^m(\theta, \phi)=Y_{\ell}^m(\pi-\theta, \pi+\phi)$ for even $\ell$ and $Y_{\ell}^m(\theta, \phi)=-Y_{\ell}^m(\pi-\theta, \pi+\phi)$ for odd $\ell$, the diffusion signal $d(\theta,\phi)$ has only even degree SH coefficients, that is, $\coeffSH=0$ for odd degree $\ell$~\cite{Caruyer:2012b,Tournier:2013,Bates:2015}. In this work, we assume that the diffusion signal is band-limited at degree $L$ such that $\coeffSH=0$ for $\ell > L$. With these considerations, we rewrite the expansion in Eq~\eqref{Eq:d_expansion} as
\begin{equation}
\label{Eq:d_expansion_trunc}
    d(\theta,\phi)=\sum_{\substack{\ell=0 \\ \ell \, \rm{even}}}^{L}\sum_{m=-{\ell}}^{\ell} \coeffSH  Y_{\ell}^m(\theta,\phi), \quad L~{\rm even}.
\end{equation}
The number of SH coefficients required to represent the diffusion signal, given in Eq~\eqref{Eq:d_expansion_trunc}, is $\nps = (L+1)(L+2)/2$~\cite{Caruyer:2012,Bates:2016}, which is the minimum number of samples attainable by any single-shell sampling scheme that allows for the accurate computation of the SHt of any band-limited antipodal signal. General spherical sampling theorems in contrast, require on the order of $2L^2$ samples~\cite{McEwen2011}.


\subsubsection*{Measurement and reconstruction}
As only a finite number of measurements of the diffusion signal can be obtained, the SHt Eq~\eqref{Eq:dcoeff} can not be solved analytically. The SH coefficients $\coeffSH$ are commonly estimated numerically using the least-squares solution of the inverse SHt Eq~\eqref{Eq:d_expansion_trunc} \cite{Caruyer:2012,Descoteaux:2007,Hess:2006,Assemlal:2009b}. Let $\diffSigVector$ denote the vector containing $M$ measurements of the diffusion signal, with $\diffSigVector= [d(\theta_0, \phi_0), d(\theta_1, \phi_1), \hdots, d(\theta_{M-1}, \phi_{M-1})  ]^T$, where $(.)^T$ denotes the transpose operator,  containing measurements taken over the sphere. The least-squares estimate is given by
\begin{equation}
     \label{Eq:ss_LS}
    \hat{\coeffSHVec} = \arg \min_{\mathbf{c}} || \mathbf{A}\mathbf{c}-\diffSigVector||^2_2 = (\mathbf{A}^H \mathbf{A})^{-1} \mathbf{A}^H \diffSigVector,
    \end{equation}
where $(.)^H$ denotes the Hermitian operator, the columns of the matrix $\mathbf{A}$ correspond to sampled versions of the even degree $\ell$ SH basis functions, $\mathbf{c} = [c_0^0,c_2^{-2},c_2^{-1},c_2^{0},c_2^{1},c_2^{2}, \hdots, c_{L}^{L} ]^T$ is the column vector of length $(L+1)(L+2)/2$ containing even $\ell$ SH coefficients of the diffusion signal and $\hat{\mathbf{c}}$ is the estimate of $\mathbf{c}$. The least-squares estimate improves with a larger $M$, as the matrix inversion tends to be ill-conditioned with a small number of samples~\cite{Khalid:2014}. Regularization of the solution is commonly used to improve the condition number of the matrix and add robustness to noise~\cite{Caruyer:2012,Descoteaux:2007}

Laplace-Beltrami regularization is widely used for reconstruction of the diffusion signal on the sphere \cite{Descoteaux:2007,Varadarajan:2015}. The Laplace-Beltrami operator $\Delta_b$ penalizes non-smooth signals \cite{Descoteaux:2007}. In \cite{Descoteaux:2007}, a measure of the deviation of a signal on the sphere $d(\theta,\phi)$ from smooth is defined as:
\begin{equation}
\label{Eq:Ef}
U(d(\theta,\phi)) =\int_{\untsph} \Big( \Delta_b d(\theta,\phi) \Big)^2 \,\sin\theta\, d\theta\, d\phi.
\end{equation}
This can be written in the spectral domain by using $\Delta_b Y^m_\ell = -\ell(\ell+1) Y^m_\ell $ and by expanding $d$ in the SH basis using Eq~\eqref{Eq:dcoeff}\cite{Descoteaux:2007},
%
\begin{equation}
\label{Eq:Ef_SH}
U(d(\theta,\phi)) = \sum_{\substack{\ell=0 \\ \ell \, \rm{even}}}^{L} \sum_{m=-\ell}^{\ell} ( \coeffSH )^2 \ell^2(\ell+1)^2
= \sum_{j=1}^{(L+1)(L+2)/2} ( c_{l(j)}^{m(j)} )^2 \ell(j)^2(\ell(j)+1)^2.
\end{equation}
$U(d(\theta,\phi))$ can also be written in matrix form as:
 \begin{equation}
 \label{Eq:Ef_SH_matrix}
U(d(\theta, \phi)) = \mathbf{c}^H \mathbf{L} \mathbf{c},
\end{equation}
where $\mathbf{L}$ is a diagonal matrix of size $(L+1)(L+2)/2 \times (L+1)(L+2)/2 $ with entries $\ell^2(\ell+1)^2$ associated with each SH degree.
The SH coefficients of the diffusion signal can then be estimated by solving the following least-squares problem with Laplace-Beltrami regularization:
\begin{equation}
     \label{Eq:reg_LS_ss}
    \hat{\mathbf{c}} = \arg \min_{\mathbf{c}}  || \mathbf{A}\mathbf{c}-\diffSigVector||^2_2 + \lambda \mathbf{c}^H \mathbf{L} \mathbf{c}
                     = (\mathbf{A}^H \mathbf{A} + \lambda\mathbf{L})^{-1} \mathbf{A}^H \diffSigVector,
    \end{equation}
where $\lambda$ is a regularization parameter. Different values of $\lambda$ are used to trade-off a smooth solution with fitting to the data~\cite{Caruyer:2012}. $\lambda$ is also necessary to improve the condition number of the matrix inversion when performing least-squares estimation for ill-conditioned problems such as when the number of measurements is not much larger than the number of coefficients to be estimated.


\subsection*{Diffusion signal in 3D and spherical polar Fourier basis}

The diffusion signal in $\mathbb{R}^3$, $d(\mathbf{q})$, can be expanded in the spherical polar Fourier (SPF) basis \cite{Assemlal:2009b}, a 3D orthonormal basis. There exist other 3D bases in diffusion MRI such as 3D-SHORE and MAP-MRI~\cite{Ozarslan:2013}, however only SPF is separable in the angular and radial directions, we use this property to define the following novel sampling scheme. The SPF expansion of $d(\mathbf{q})$ is given by,
\begin{equation}
\label{Eq:E_expansion}
    d(\mathbf{q})=\sum_{n=0}^{N}\sum_{\substack{\ell=0 \\ \ell \, \rm{even}}}^{L}\sum_{m=-{\ell}}^{\ell} \coeffSPF  R_n(q)Y_{\ell}^m(\theta,\phi),  \quad L~{\rm even},
\end{equation}
where $q = |\mathbf{q}|$ and the radial functions $R_n$ are Gaussian-Laguerre polynomials with
\begin{equation}
\label{Eq:radial_func}
  R_n(q)=\bigg[\frac{2}{\zeta^{1.5}}\frac{n!}{\Gamma(n+1.5)}\bigg]^{0.5}\exp\bigg(\frac{-q^2}{2\zeta}\bigg)L_n^{1/2}\bigg(\frac{q^2}{\zeta}\bigg),
\end{equation}
where $\zeta$ denotes the scale factor and $L_n^{1/2}$ are the $n$-th generalized Laguerre polynomials of order half. The expansion coefficients are given by
\begin{equation}\label{Eq:Ecoeff}
\coeffSPF = \langle d(\mathbf{q}),  R_n(q)Y_{\ell}^m(\theta,\phi)\rangle 
              = \int_{q} \intsph d(\mathbf{q})R_n(q)\conj{Y_{\ell}^m(\theta,\phi)}\, q^2 \sin(\theta)\, d\theta\, d\phi\, dq.
\end{equation}
The expansion Eq~\eqref{Eq:E_expansion} is band-limited at radial order $N$ and angular order $L$.


\subsubsection*{Measurement and reconstruction}
The SPF coefficients are typically computed numerically using the least-squares solution to Eq~\eqref{Eq:E_expansion} \cite{Assemlal:2009b}. Let the length $M$ measurement vector $\diffSigVector$ contain measurements of the diffusion signal taken on a multi-shell sampling scheme with $\diffSigVector= [d(\mathbf{q}_0), d(\mathbf{q}_1), \hdots, d(\mathbf{q}_{M-1})  ]^T$. The least-squares estimate is given by
\begin{equation}
     \label{ms_LS}
    \hat{\coeffSPFVec} = \arg \min_{\coeffSPFVec} || \mathbf{B}\coeffSPFVec-\diffSigVector||^2_2
                     = (\mathbf{B}^H \mathbf{B})^{-1} \mathbf{B}^H \diffSigVector,
    \end{equation}
where the columns of the matrix $\mathbf{B}$ correspond to sampled versions of the SPF functions, $\coeffSPFVec = [(e)_{000},(e)_{02-2}, \hdots, (e)_{N L L} ]^T$ is the vector containing even angular degree $\ell$ SPF coefficients of the diffusion signal and $\hat{\coeffSPFVec}$ is the estimate of $\coeffSPFVec$. Laplace-Beltrami regularization in the angular direction combined with a radial regularization term which penalizes higher radial degrees is commonly used in multi-shell reconstruction~\cite{Assemlal:2009b,Merlet:2013}.
Using this regularization, the SPF coefficients of the diffusion signal can be estimated as follows:
%
%
\begin{equation}
     \label{Eq:reg_LS_ms}
    \hat{\coeffSPFVec} = \arg \min_{\coeffSPFVec} || \mathbf{B}\coeffSPFVec-\diffSigVector||^2_2 + \lambda_\ell \coeffSPFVec^H \mathbf{L} \coeffSPFVec + \lambda_n \coeffSPFVec^H \mathbf{N} \coeffSPFVec
                     = (\mathbf{B}^H \mathbf{B} + \lambda_\ell\mathbf{L} + \lambda_n\mathbf{N} )^{-1} \mathbf{B}^H \diffSigVector,
    \end{equation}
where $\mathbf{N}$ is a diagonal matrix of size $(N+1)(L+1)(L+2)/2 \times (N+1)(L+1)(L+2)/2 $ with entries $n^2(n+1)^2$ associated with each SPF radial degree, $\mathbf{L}$ is a matrix of size $(N+1)(L+1)(L+2)/2\times (N+1)(L+1)(L+2)/2 $ with entries $\ell^2(\ell+1)^2$ associated with each SPF angular degree along the diagonal and $\lambda_\ell$ and $\lambda_n$ are the angular and radial regularization parameters respectively.


\subsection*{Rician and non-central Chi distributed noise}\label{Sec:noise} 
In diffusion MRI, the diffusion signal typically has a Rician or non-central Chi distribution (NCC)~\cite{Varadarajan:2015}.  The negative log-likelihood for a NCC distribution given by~\cite{Varadarajan:2015,Canales:2015},

\begin{equation}\label{Eq:NCC_NLL}
\mathcal{L}_{\rm ncc}(\diffSigVector|\coeffGenVec) = \sum_{p=1}^{M} \Bigg\{ \frac{[\basisGen(\coeffGenVec)]^2_p}{2\sigma^2} - \ln\Bigg( \frac{ I_{C-1} \Big( \frac{[\basisGen(\coeffGenVec)]_p [\diffSigVector]_p}{\sigma^2} \Big) }{[\basisGen(\coeffGenVec)]^{C-1}_p} \Bigg) \Bigg\},
\end{equation}
where $\sigma^2$ is the variance of the real and imaginary parts of the Gaussian noise of the complex diffusion MRI signal, $C$ is the number of coils, $\basisGen$ is the operator mapping $\coeffGenVec$ to $\diffSigVector$ and $\coeffGenVec$ is the vector of coefficients of the diffusion signal in that basis. Eq~\eqref{Eq:NCC_NLL} reduces to the Rician negative log-likelihood when $C=1$.

In \cite{Varadarajan:2015}, a method is proposed for solving the penalized maximum likelihood~(PML) estimator for the NCC distribution:
\begin{equation}\label{Eq:PML}
\hat{\coeffGenVec} = \arg \min_{\coeffGenVec} \mathcal{L}_{\rm ncc}(\diffSigVector|\coeffGenVec) + R(\coeffGenVec),
\end{equation}
where $R(\coeffGenVec)$ is some regularization term applied to $\coeffGenVec$.
The algorithm in~\cite{Varadarajan:2015}, when applied to estimating the coefficients of the diffusion signal in some orthonormal basis when the measurements are corrupted by NCC or Rician noise, consists of the following steps to find estimate of the coefficients $\hat{\coeffGenVec}$ :
\begin{enumerate}
\item Initialize ($\hat{\coeffGenVec})_0$.
\item Until convergence:
     \begin{equation}
     \label{Eq:MM}
    (\hat{\coeffGenVec})_{i+1} = \arg \min_{\coeffGenVec: \basisGen(\coeffGenVec)\geq 0 } \frac{1}{2\sigma^2} || \basisGen(\coeffGenVec)-(\tilde{\diffSigVector})_i||^2_2 + R(\coeffGenVec),
    \end{equation}
 where
     \begin{equation}
     \label{Eq:y_update}
    [(\tilde{\diffSigVector})_i]_p = [\diffSigVector]_p \frac{I_{C}\Big(\frac{ [\basisGen((\hat{\coeffGenVec})_i)]_p [\diffSigVector]_p}{\sigma^2} \Big)}{I_{C-1}\Big(\frac{[\basisGen((\hat{\coeffGenVec})_i)]_p [\diffSigVector]_p}{\sigma^2} \Big)}.
    \end{equation}
\end{enumerate}
 When $\basisGen((\hat{\coeffGenVec})_{i+1})\geq 0 $ the solution of Eq~\eqref{Eq:MM} is simply regularized least-squares, for example Eq~\eqref{Eq:reg_LS_ss} or Eq~\eqref{Eq:reg_LS_ms}. When non-negativity does not hold, Eq~\eqref{Eq:MM} is solved as a quadratic program. We note that this algorithm is similar to the multiplicative Richardson-Lucy algorithm for estimating the fiber orientation distribution function termed RUMBA-SD in~\cite{Canales:2015}.

 Eq~\eqref{Eq:y_update} requires an estimate of $\sigma^2$, which is difficult to obtain accurately from real data. Hence in RUMBA-SD $\sigma^2$ is also estimated iteratively. We incorporate this into the above denoising method, where the following estimate $\hat{\sigma}^2$  is obtained by minimizing the negative log-likelihood function, given in Eq~\eqref{Eq:NCC_NLL}, with respect to $\sigma^2$:
%
  \begin{equation}
    \label{Eq:sigma_update}
    (\hat{\sigma}^2)_{i} = \frac{1}{CM}\Bigg( \frac{\diffSigVector^H\diffSigVector + \basisGen((\hat{\coeffGenVec})_i)^H \basisGen((\hat{\coeffGenVec})_i)}{2} - \sum_{p=0}^{M-1} [\diffSigVector]_p [\basisGen((\hat{\coeffGenVec})_i)]_p \frac{I_{C}\Big(\frac{ [\basisGen((\hat{\coeffGenVec})_i)]_p [\diffSigVector]_p}{(\hat{\sigma}^2)_{i-1}} \Big)}{I_{C-1}\Big(\frac{[\basisGen((\hat{\coeffGenVec})_i)]_p [\diffSigVector]_p}{(\hat{\sigma}^2)_{i-1}} \Big)} \Bigg) .
   \end{equation}
%
\subsection*{Gaussian mixture model of the diffusion signal} \label{Sec:GMM}
The Gaussian mixture model~(GMM) is commonly used to simulate the diffusion signal in white-matter tissue and can be used to represent fiber crossings as well as single fibers~\cite{Assemlal:2009b,Daducci:2014}. The diffusion weighted signal for a single voxel is given by
\begin{equation} \label{Eq:diff_model}
    d(\bv{q})=d(\theta,\phi; b)=\sum_{k=1}^{F} f_{k}e^{-b \bv{u}(\theta, \phi)^{T} \bv{D}_{k} \bv{u}(\theta, \phi)},
\end{equation}
where $F$ is the number of fiber populations, $f_{k}$ are the corresponding fractions such that $\sum_{k=1}^{F} f_{k} = 1$, $\bv{D}_{k}$ encodes the diffusivity properties for the $k$th fiber in the voxel and the diffusion weighting is given by the $b$-value, where $b$ is proportional to $q^2$. Each fiber's tensor is computed from a rotated version of a tensor, $\bv{D}={\rm diag}(\lambda_{1}, \lambda_{2}, \lambda_{3})$, with
$\bv{D}_{k} = \bv{R}_{k}^T\bv{D}\bv{R}_{k}$, where $\lambda_{1}$ is the diffusivity along the main axis of a fiber while $\lambda_{2}$ and $\lambda_{3}$ are the diffusivities in the plane perpendicular to it, and $\bv{R}_{k}$ is the rotation matrix that rotates the $k$th fiber to the direction of the $k$th fiber population.

In the numerical experiments where the effect of noise is considered, we add Rician noise to the diffusion weighted signal as~\cite{Gudbjartsson:1995}
\begin{equation} \label{Eq:noisy_signal}
    d(\theta,\phi; b)_{\rm{noisy}}= \sqrt{ \big(d(\theta,\phi; b) +\eta_{1} \big)^{2} +\eta_{2}^2},
\end{equation}
with $\eta_{1},\eta_{2} \sim \mathcal{N}(0,\sigma^{2})$ and $\sigma = d_{0}/{\mathrm{SNR}}$. The signal-to-noise ratio~(SNR) controls the level of noise on the baseline image with $b=0$, $d_0$~\cite{Daducci:2014}.


\subsection*{Spherical harmonic band-limit selection for each shell}\label{Sec:SH_L_per_shell}

The SH band-limit $L$ required to accurately represent the diffusion signal depends on the $b$-value~\cite{Daducci:2011,Hess:2006,Tournier:2013}. In~\cite{Daducci:2011}, a study is carried out, using the GMM of the diffusion signal to determine what $L$ is required for different $b$-values.

\begin{figure}[!h]
\centering
  \includegraphics[trim=40 10 40 20, clip, width=0.48\textwidth]{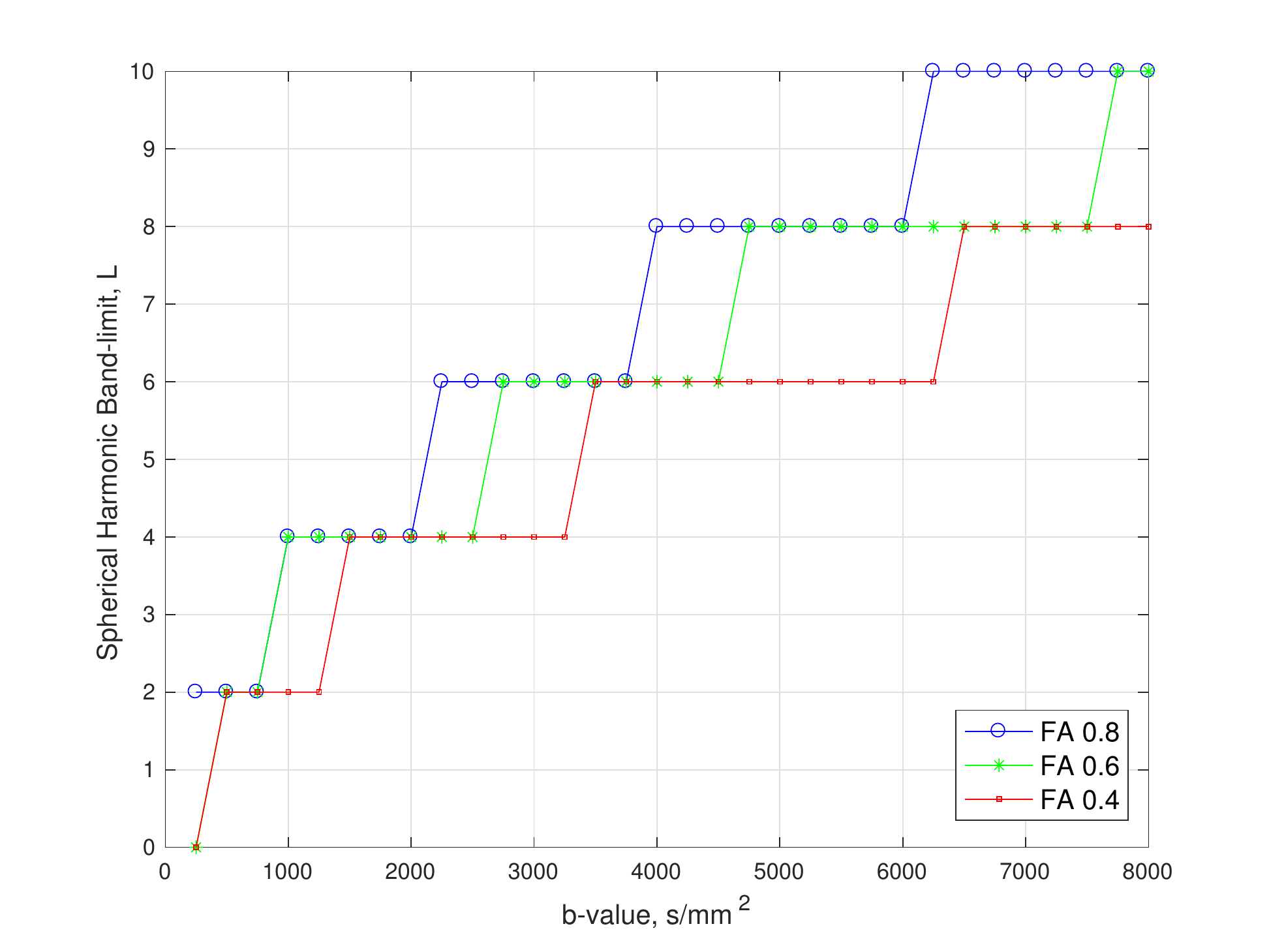} 
  \caption{{\bf Spherical harmonic band-limit for different $b$-values.} The SH band-limit $L$ required to accurately represent the diffusion signal at different $b$-values. Results are reported for three levels of anisotropy of the signal, i.e. fractional anisotropies (FA) $0.4$, $0.6$ and $0.8$.}
  \label{fig:SH_bandlimit}
\end{figure}

The diffusivities $\lambda_{1}$, $\lambda_{2}$ and $\lambda_{3}$ were varied in this experiment to produce different fractional anisotropies (FA), where FA $\in [0,1]$ is a widely-used measure of the anisotropy of diffusion, for various numbers of fiber populations and, various volume fractions and crossing angles. The band-limit is found to increase with the FA and be independent of the other factors due to the GMM being the sum of the signals from the different fiber populations. We repeat this experiment and confirm the results shown in~\figref{fig:SH_bandlimit} for FA $=0.4,0.6$ and 0.8. \figref{fig:SH_bandlimit} is used for determining what $L$ to use in each shell of the proposed single and multi-shell schemes presented later in the paper.

\section*{Proposed single-shell sampling scheme} \label{Sec:ss}
The proposed single-shell sampling grid and SHt were presented in~\cite{Bates:2016} and are summarized here so that the paper is self-contained. The sampling grid structure enables the development of a novel SHt which, unlike least-squares, has well-conditioned matrices when the number of samples is equal to the number of SH coefficients.  We also extend the sampling scheme so that it allows for more robust reconstruction by using regularization and considering the non-Gaussian nature of the noise.


\subsection*{Sampling grid and transform}\label{Sec:ss_sampling_scheme}
\begin{figure}[!h]
 \centering
  \subfloat[]{
  \includegraphics[width=0.2\textwidth]{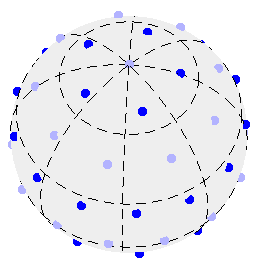}} \hspace{15mm}
   \subfloat[]{
  \includegraphics[width=0.2\textwidth]{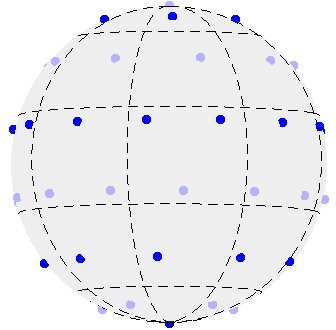}}
  \caption{{\bf Single-shell sampling grid.} Proposed single-shell sampling grid for $L=6$ (a) North pole view and (b) side view.}
  \label{fig:ss_sampling_grid}
\end{figure}

In \cite{Bates:2016}, an iso-latitude sampling scheme is presented with $(L+2)/2$ iso-latitude rings, located at
\begin{align}\label{Eq:theta_proposed}
\bv\theta \dfn [\theta_0, \theta_1, \hdots,\, \theta_{L/2}]^T,\quad L \,\,\textrm{even},
\end{align}
and sample equally spaced along longitude, with $k$-th sample location, denoted by $\phi^j_k$, in the ring placed at $\theta_j$ given by
\begin{align}\label{Eq:phi_proposed}
\phi^j_k \dfn \frac{2k\pi}{4j+1},\quad j \in[0,\, L/2], \quad k \in[0,\, 4j].
\end{align}
The SH band-limit $L$ is chosen, depending on the $b$-value of the shell, using~\figref{fig:SH_bandlimit}. \figref{fig:ss_sampling_grid} shows the proposed single-shell sampling scheme for band-limit $L=6$ with measurements taken on both hemispheres; antipodal points are shown lighter in color. Measurements can either be taken at either a point or its antipodal point, thus measurements can be taken on one hemisphere or spread between both hemispheres. 

It can be seen in~\figref{fig:ss_sampling_grid} that the proposed sampling scheme is not uniform by design but neither does it have dense sampling on any part of the sphere. 
\begin{remark}[Dimensionality of proposed single-shell scheme]
The total number of samples in the proposed scheme is
\begin{equation}\label{Eq:number_samples_antipodal}
\nps = \sum_{j=0}^{L/2}(4j+1) = \frac{(L+1)(L+2)}{2}.
\end{equation}
Hence the number of samples is equal to the number of coefficients in the SH basis.
\end{remark}


\subsubsection*{Novel spherical harmonic transform} \label{Sec:ss_novel_SHT}
The sampling grid, defined by the vectors $\bv\theta$ and $\phi^j_k$, given in Eq~\eqref{Eq:theta_proposed} and Eq~\eqref{Eq:phi_proposed} enables the following novel SHt algorithm. That is, the isolatitude rings of samples enables the separation of the transform in $\theta$ and $\phi$, as well as the SH coefficients to be calculated one order $m$ at a time.  Let $\bv\theta^{m} \dfn [\theta_{|m/2|},\, \theta_{|m/2|+1},\,\hdots,\,\theta_{L/2}]^T \subset\bv{\theta}, \quad |m| \leq L, \, m~\rm{even}$ and $\bv\theta^{m} \dfn \bv\theta^{m+1}, \quad m~\rm{odd}$.
The vector $\mathbf{g}_m \equiv G_m(\bv\theta^{m})$, with
\begin{equation}\label{Eq:gm_integral}
G_m(\theta_j) \dfn \intphi \! d(\theta_j,\phi) e^{-im\phi} d\phi = 2\pi \sum_{\substack{\ell=|m| \\ \ell \, \rm{even}}}^{L} \coeffSH  \plms_\ell^m(\theta_{j}),
\end{equation}
is defined for $|m|\leq L$ and $\theta_j\in \bv{\theta}$, where $\plms_\ell^m(\theta_{j}) \dfn Y_\ell^m(\theta_{j},0)$. The SH coefficients of order $m$ can be recovered from Eq~\eqref{Eq:gm_integral} by setting up a system of linear equations of size $\lceil(L+1-|m|)/2\rceil$, given by
\begin{equation}\label{Eq:gtof}
\mathbf{g}_m  = \mathbf{P}_m \mathbf{c}_m,\, \quad |m|\leq L,
\end{equation}
where
\begin{equation}
\mathbf{c}_m  = \begin{cases}\big[ \shc{c}{{|m|}}{m},\,  \shc{c}{{|m|}+2}{m},\,\hdots,\, \shc{c}{L}{m}\big]^T, \quad m~\rm{even},\\
\big[ \shc{c}{{|m|+1}}{m},\,  \shc{c}{{|m|}+3}{m},\,\hdots,\, \shc{c}{L}{m}\big]^T, \quad m~\rm{odd},
\end{cases}
\end{equation}
and $\mathbf{P}_m$ is defined as
\begin{equation*}\label{Eq:P_matrix_even}
\mathbf{P}_m \dfn \begin{small}\setlength{\arraycolsep}{1mm}
2\pi\begin{pmatrix}
   \plms_{|m|}^m(\theta_{s}) & \plms_{{|m|}+2}^m(\theta_{s}) & \cdots & \plms_{L}^m(\theta_{s}) \\[2mm]
   \plms_{|m|}^m(\theta_{s+1}) & \plms_{{|m|}+2}^m(\theta_{{s}+1}) & \cdots & \plms_{L}^m(\theta_{s+1}) \\
   \vdots  & \vdots  & \ddots & \vdots  \\
   \plms_{|m|}^m(\theta_{\frac{L}{2}}) & \plms_{{|m|}+2}^m(\theta_{\frac{L}{2}}) & \cdots & \plms_{L}^m(\theta_{\frac{L}{2}})\\
  \end{pmatrix}\end{small},
\end{equation*}
for even $m$ and
\begin{equation*}\label{Eq:P_matrix_odd}
\mathbf{P}_m \dfn \begin{small}\setlength{\arraycolsep}{0.05mm}
2\pi\begin{pmatrix}
   \plms_{|m|+1}^m(\theta_{s}) & \plms_{{|m|}+3}^m(\theta_{s}) & \cdots & \plms_{L}^m(\theta_{s}) \\[2mm]
   \plms_{|m|+1}^m(\theta_{s+1}) & \plms_{{|m|}+3}^m(\theta_{s+1}) & \cdots & \plms_{L}^m(\theta_{s+1}) \\
   \vdots  & \vdots  & \ddots & \vdots  \\
   \plms_{|m|+1}^m(\theta_{\frac{L}{2}}) & \plms_{{|m|}+3}^m(\theta_{\frac{L}{2}}) & \cdots & \plms_{L}^m(\theta_{\frac{L}{2}})\\
  \end{pmatrix}\end{small},
\end{equation*}
for odd $m$. Here $s = \lceil{|m|}/{2}\rceil$, where $\lceil\cdot \rceil$ denotes the integer ceiling function.

 The integral in Eq~\eqref{Eq:gm_integral} can be accurately evaluated as a summation provided there are at least $2m+1$ samples along $\phi$ \cite{Khalid:2014}. The SH coefficient $\coeffSH$ are computed from higher to lower order $m$. By exploiting the orthogonality of the SH functions over order $m$, the contribution from SH with higher $m$ can be removed from the diffusion signal. This means that less samples are required to calculate $\coeffSH$ with smaller $m$, which enables the proposed sampling scheme to obtain the number of samples equal to the number of SH coefficients, see~\cite{Bates:2015,Khalid:2014} for further details.  In order to accurately compute the SHt, we choose the sampling points along co-latitude such that the matrix $\mathbf{P}_m$ is well-conditioned for each $|m|\leq L$~\cite{Khalid:2014,Bates:2015}. The $\mathbf{P}_m$ matrices are of size $\lceil(L+1-|m|)/2\rceil \times \lceil(L+1-|m|)/2\rceil$, with the largest $\mathbf{P}_0 = (L/2 +1)^2$, compared with the least-squares matrix $\mathbf{A}$ in Eq~\eqref{Eq:ss_LS} and Eq~\eqref{Eq:reg_LS_ss} which is size $L(L+1)/2 \times L(L+1)/2 $. The novel SHt allows for accurate reconstruction, with the reconstruction error on the order of machine precision for antipodally symmetric noise-free signals band-limited in the SH basis. In reality the diffusion signal is corrupted by noise and it is necessary to regularize the solution. 


\subsection*{Regularization} \label{Sec:ss_reg}
Here we develop regularization for the sampling scheme presented above. Equation Eq~\eqref{Eq:Ef_SH} can be rewritten with the order of summation changed to be an outer summation over order $m$ and an inner summation over degree $\ell$ as,
 \begin{equation}
 \label{Eq:Ef_SH_sum_m}
U(d(\theta,\phi)) =\sum_{m=-L}^{L} \sum_{\substack{\ell=0 \\ \ell \, \rm{even}}}^{L} ( \coeffSH )^2 \ell^2(\ell+1)^2.
\end{equation}
We define $U(d(\theta,\phi),m)$ as the per order measure of deviation of a signal on the sphere from smooth with,
 \begin{equation}
 \label{Eq:Ef_SH_m}
U(d(\theta,\phi),m) =\sum_{\substack{\ell=0 \\ \ell \, \rm{even}}}^{L} ( \coeffSH )^2 \ell^2(\ell+1)^2 =  \mathbf{c}_m \mathbf{L}_m  \mathbf{c}_m,
\end{equation}
where $\mathbf{L}_m = \rm{diag}\Big( \ell^2(\ell+1)^2, \; \ell =m:L, \; \ell \, \rm{even}\Big)$.
 Using the novel SHt presented above, the regularized least-squares method of obtaining the SH coefficients Eq~\eqref{Eq:reg_LS_ss} can be solved equivalently for each $m$ as:
      \begin{equation} \label{Eq:reg_nsht}
    \hat{\mathbf{c}}_m = \arg \min_{\mathbf{c}_m} || \mathbf{P}_m\mathbf{c}_m-\mathbf{g}_m||^2_2 + \lambda \mathbf{c}_m^H \mathbf{L}_m \mathbf{c}_m 
                       = (\mathbf{P}_m^H \mathbf{P}_m + \lambda\mathbf{L}_m)^{-1} \mathbf{P}_m^H \mathbf{g}_m,
    \end{equation}
resulting in a Laplace-Beltrami regularized novel SHt.


\subsection*{Non-Gaussian noise removal}\label{Sec:ss_denoising}

We further extend the proposed sampling scheme presented above to include non-Gaussian noise removal by incorporating it into the majorize-minimize framework for denoising diffusion MRI with NCC distributed or Rician noise~\cite{Varadarajan:2015}. The steps for the resulting reconstruction algorithm are as follows:
\begin{enumerate}
\item Initialize $(\hat{\mathbf{c}})_0$ using the regularized novel SHt Eq~\eqref{Eq:reg_nsht}.
\item Until convergence, calculate the SH coefficients of the diffusion signal estimate $\hat{\mathbf{c}}$ for each order $m$:
     \begin{align}
     \label{Eq:nsht_MM}
    (\hat{\mathbf{c}}_m)_{i+1} &= \arg \min_{\mathbf{c}_m} \frac{1}{2\sigma^2} || \mathbf{P}_m \mathbf{c}_m-(\tilde{\mathbf{g}}_m)_i||^2_2 + \lambda \mathbf{c}_m^H \mathbf{L}_m \mathbf{c}_m,\nonumber \\
    &= (\mathbf{P}_m^H \mathbf{P}_m + \lambda\mathbf{L}_m)^{-1} \mathbf{P}_m^H (\tilde{\mathbf{g}}_m)_i.
    \end{align}
    where $(\tilde{\mathbf{g}}_m)_i$ is calculated using Eq~\eqref{Eq:gm_integral} with $\diffSigVector$ replaced by $(\tilde{\diffSigVector})_i$ and $(\tilde{\diffSigVector})_i$ is calculated using Eq~\eqref{Eq:y_update}.  In Eq~\eqref{Eq:y_update} an updated estimate of $\sigma^2$ is used calculated using Eq~\eqref{Eq:sigma_update}. 

\end{enumerate}


\section*{Proposed multi-shell sampling scheme} \label{Sec:ms}
We here propose a novel multi-shell sampling scheme which is a generalization of the single-shell scheme.

\subsection*{Sampling scheme and novel SPF transform} \label{sec:ms_sampling_scheme}
Due to the separability of the SPF basis, the 3D transform for calculating the diffusion signal coefficients Eq~\eqref{Eq:Ecoeff} can be separated into transforms in the radial and angular dimensions. Rearranging Eq~\eqref{Eq:Ecoeff},
\begin{equation}\label{Eq:Ecoeff_sep}
\coeffSPF = \int_{q} R_n(q) \, q^2 \intsph d(\mathbf{q}) \conj{Y_{\ell}^m(\theta,\phi)}\, \sin(\theta)\, d\theta\, d\phi\, dq,
\end{equation}
where the inner integral is the SHt Eq~\eqref{Eq:dcoeff}. We use the novel SHt presented in the Proposed Single-shell Sampling Scheme Section to perform the angular transform.

\begin{figure}[!h]
  \centering
  \subfloat[]{
  \includegraphics[width=0.22\textwidth]{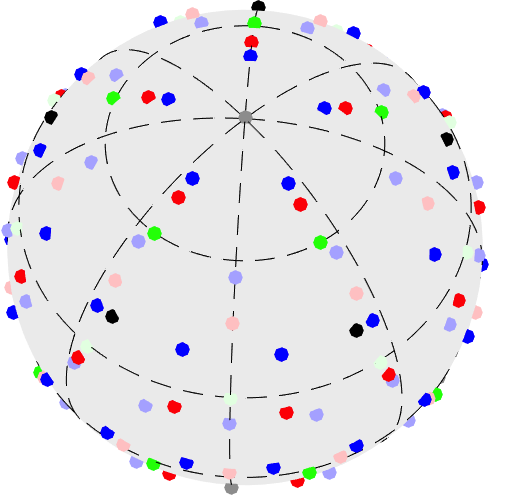}} \hspace{15mm}
   \subfloat[]{
  \includegraphics[width=0.22\textwidth]{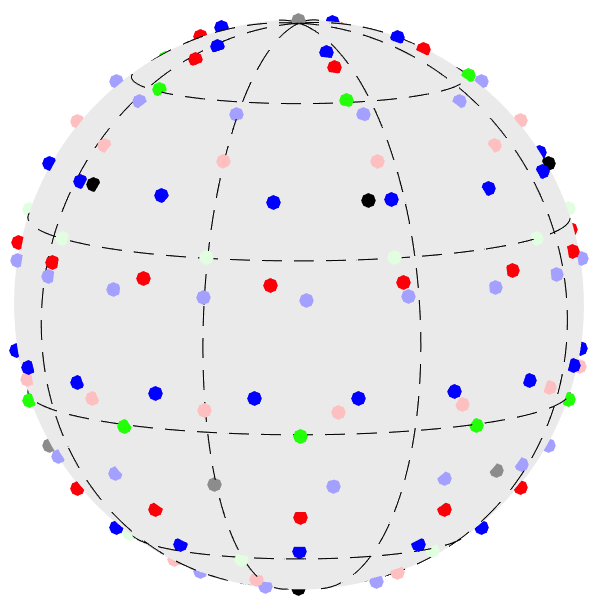}}
  \caption{{\bf Multi-shell sampling grid projected onto a sphere.} Proposed multi-shell sampling scheme for $N=3$ and $L(s)=[2,4,6,8]$ (a) North pole view and (b) side view.}
  \label{fig:sampling_grid}
\end{figure}

For the radial transformation, Gauss-Laguerre quadrature can be used, where $N+1$ sampling nodes is sufficient for exact quadrature~\cite{Abramowitz:1964,Leistedt:2012}, enabling Eq~\eqref{Eq:Ecoeff_sep} to be written as,
\begin{equation}\label{Eq:Ecoeff_GL_quad}
\coeffSPF = \sum_{s=0}^{N} w_s R_n(q_s) \intsph d(\mathbf{q_s}) \conj{Y_{\ell}^m(\theta,\phi)}\, \sin(\theta)\, d\theta\, d\phi,
\end{equation}
where the $q^2$ in Eq~\eqref{Eq:Ecoeff_sep}  is incorporated into the weights $w_s$ and $s$ is the shell index. 
The $N+1$ shells of the proposed multi-shell sampling scheme are placed at $q_s = \sqrt{\zeta x_s}$ where $x_s$ are the roots of the $(N+1)$-th generalized Laguerre function of order a half. We determine the corresponding weights to be
\begin{equation}
\label{Eq:weight}
w_s =\frac{0.5\zeta^{0.5}\Gamma(N+2.5)x_se^{x_s}}{(N+1)!(N+2)^2[L_{N+2}^{0.5}(x_s)]^2}.
\end{equation}
 We set the scaling factor $\zeta$ so that shells are located at $b$-values within an interval of interest. For sampling within each shell, we use the proposed single-shell sampling scheme presented in the Proposed Single-shell Sampling Scheme Section. The SH band-limit, and therefore the number of samples in each shell, is determined using \figref{fig:SH_bandlimit}. 
\begin{remark}[Dimensionality of proposed multi-shell scheme]
The total number of samples in the proposed scheme is
\begin{equation}\label{Eq:number_samples_antipodal}
\sum_{s=0}^{N} \frac{(L(s)+1)(L(s)+2)}{2},
\end{equation}
where $L(s)$ is a vector of length $N+1$ containing the SH band-limit for each shell. That is, the multi-shell sampling scheme has the number of samples in each shell equal to the number of SH coefficients in each shell $\nps$ Eq~\eqref{Eq:number_samples_antipodal}. This is enabled through the novel SPF transform (SPFt) which uses the separability of the SPF basis to perform the angular and radial transforms separately. This means that a different SH band-limit can be used in each shell. To our knowledge no other reconstruction algorithm in diffusion MRI enables this; standard least-squares calculates all the SPF coefficients in one step using Eq~\eqref{Eq:reg_LS_ms} and so must use a single spherical harmonic band-limit for all shells.   
\end{remark}
\figref{fig:sampling_grid} shows the proposed sampling scheme with $N=3$ and $L(s)=[2,4,6,8]$ projected onto a single sphere, samples on the inner most to outer most shell are shown in black, green, red and blue for each shell respectively. Locations where antipodal symmetry is used to infer the value of the signal are lighter in color.


\subsection*{Regularization}\label{Sec:ms_reg}
Here we increase the robustness of the novel SPFt presented above by regularizing the solution. This is done by
extending the single-shell regularization to multi-shell reconstruction by adding a radial regularization term which penalizes higher radial degrees as is done in the standard regularized least-squares method Eq~\eqref{Eq:reg_LS_ms}.

The regularized penalized least-squares method of computing the SPF coefficients Eq~\eqref{Eq:reg_LS_ms} can be solved equivalently for each radial degree $n$ and for each shell by calculating the SH coefficients of order $m$ as:
    \begin{align}
     \label{Eq:reg_nsht_ms}
    (\hat{\mathbf{c}}_m)_{ns} &= \arg \min_{\SHorderVecPerShellAndn} ||\mathbf{P}_m\SHorderVecPerShellAndn-\mathbf{g}_m||^2_2 +\lambda_\ell {\SHorderVecPerShellAndn}^H \mathbf{L}_m \SHorderVecPerShellAndn \nonumber \\
    & +\lambda_n \SHorderVecPerShellAndn^H \mathbf{N}_n \SHorderVecPerShellAndn
                       = (\mathbf{P}_m^H \mathbf{P}_m + \lambda_\ell\mathbf{L}_m+\lambda_n\mathbf{N}_n)^{-1} \mathbf{P}_m^H \mathbf{g}_m,
    \end{align} 
where $\mathbf{N}_n$ has entries $n^2(n+1)^2$ along the diagonal depending on what iteration $n$ of the radial order. Let $(\hat{c}^m_\ell)_{n s}$ denote the SH coefficient calculated for shell $s$ and radial order $n$. The SPFt in Eq~\eqref{Eq:Ecoeff_GL_quad} can then be computed with radial and angular regularization as
\begin{equation}\label{Eq:ms_reg_trans}
\hat{e}_{n\ell m}= \sum_{s=0}^{N} w_s R_n(q_s)(\hat{c}^m_\ell)_{n s}.
\end{equation}

\begin{remark}[Conditioning of matrix inversion]\label{Re:conditioning}
The $\mathbf{P}_m$ matrices are designed to be well-conditioned. For instance, for the multi-shell sampling scheme with parameters $N$ and $L(s)$ as shown in~\figref{fig:sampling_grid}, the maximum condition number of the $\mathbf{P}_m$ matrices is 17, whereas the least-squares matrix $\mathbf{B}$ in Eq~\eqref{ms_LS} is ill-conditioned when the number of samples used in each shell is equal to the number of SH coefficients, with a condition number of $1.5 \times 10^{18}$ without regularization. Though regularization improves this, the condition number is at least on the order of $10^{13}$ for all values of the regularization parameters $\lambda_\ell, \, \lambda_n \in [0,1]$.
\end{remark}


\subsection*{Non-Gaussian noise removal}\label{Sec:ms_denoising}
We further improve the robustness of the novel SPFt with regularization by including non-Gaussian noise removal by extending the proposed single-shell denoising algorithm to multi-shell as follows:
\begin{enumerate}
\item Initialize $(\hat{\coeffSPFVec})_0$ using the regularized novel SPFt Eq~\eqref{Eq:ms_reg_trans}.
\item Until converge of the SPF coefficient estimates $ \hat{\coeffSPFVec}$,  for each shell $s$ and radial degree $n$,  for each order $m$:
  \begin{align}
     \label{Eq:SPF_MM}
 &((\hat{\mathbf{c}}_m)_{ns})_{i+1} = \arg \min_{\SHorderVecPerShellAndn} \frac{1}{2\sigma^2} || \mathbf{P}_m \SHorderVecPerShellAndn-(\tilde{\mathbf{g}}_m)_i||^2_2
 + \lambda_\ell \SHorderVecPerShellAndn^H \mathbf{L}_m \SHorderVecPerShellAndn   \nonumber \\
  & + \lambda_n \SHorderVecPerShellAndn^H \mathbf{N}_n \SHorderVecPerShellAndn =(\mathbf{P}_m^H \mathbf{P}_m + \lambda_\ell \mathbf{L}_m  + \lambda_n \mathbf{N}_n)^{-1} \mathbf{P}_m^H (\tilde{\mathbf{g}}_m)_i,
    \end{align}
    where $(\tilde{\mathbf{g}}_m)_i$ is calculated using Eq~\eqref{Eq:gm_integral} with $\diffSigVector$ replaced by $(\tilde{\diffSigVector})_i$ and $(\tilde{\diffSigVector})_i$ is calculated using Eq~\eqref{Eq:y_update}. In Eq~\eqref{Eq:y_update} an updated estimate of $\sigma^2$, is used calculated using Eq~\eqref{Eq:sigma_update}. The SPF coefficients $ (\hat{\coeffSPFVec})_{i+1}$ are obtained from $((\hat{\mathbf{c}}_m)_{ns})_{i+1}$ using Eq~\eqref{Eq:ms_reg_trans}.
\end{enumerate}


\section*{Validation}\label{Sec:Validation}
We validate the proposed single and multi-shell sampling schemes using synthetic and real diffusion MRI data-sets. This is done by comparing the proposed schemes to the standard regularized least-squares method of reconstruction given by Eq~\eqref{Eq:reg_LS_ss} and Eq~\eqref{Eq:reg_LS_ms} which we denote LS-Reg and least-squares used in the PML estimation method presented in the Rician and Non-central Chi Distributed Noise Subsection to account for non-Gaussian noise, denoted by LS-Reg-Denoised. The comparison is done using the number of samples, using the single-shell and multi-shell sampling grids presented in the Proposed Single-shell Sampling Scheme Section and the Proposed Multi-shell Sampling Scheme Section respectively. Thus, we are comparing both how using the novel SHt, for the single-shell scheme, and the novel SPFt, for the multi-shell scheme, rather than using least-squares transforms and how taking into account the non-Gaussian noise changes the reconstruction accuracy.


\subsection*{Synthetic data}
We quantitatively validate the proposed single and multi-shell sampling schemes using synthetic data-sets using the GMM, presented in the Gaussian Mixture Model of the Diffusion Signal Section, with Rician noise added.

The first synthetic data-set is generated to test the effect of crossing angle on reconstruction accuracy. The GMM with two fibers of equal volume fractions ($f_1 = f_2 =0.5$) and diffusivities $\lambda_1=1.7 \times 10^{-3} \rm{mm^2/s}$,  $\lambda_2 = \lambda_3 = 0.3 \times 10^{-3} \rm{mm^2/s}$ is used to simulate the diffusion signal in a voxel. The results are averaged over 100 different noise realizations for SNR $= 10, 20$ and $30$, corresponding to a low, moderate and high SNR, and the angle between the fibers is varied from $0^\circ$ to $90^\circ$.  

Another synthetic data-set is generated to determine how FA changes reconstruction accuracy. Using a single fiber volume fractions ($f_1 = 1$), the FA is varied from $0.5$ to $1$ and the results are averaged over 100 different noise realizations for SNR $= 10, 20$ and $30$. 


\subsubsection*{Single-shell}\label{Sec:val_ss}
Here we evaluate the proposed single-shell scheme with novel regularized SHt presented in the Proposed Single-shell Sampling Scheme - Regularization Subsection, denoted nSHt-Reg, and also the proposed scheme additionally modeling the non-Gaussian noise presented in Proposed Single-shell Sampling Scheme - Non-Gaussian Noise Removal Subsection, denoted nSHt-Reg-Denoised. As evaluation metrics, we use the normalized root mean-squared error (NRMSE) of the estimated SH coefficients given by,
\begin{equation}
\label{eq:SH_error}
\rm{NRMSE}_{\coeffSHVec} = \frac{||\hat{\coeffSHVec}-\coeffSHVec||_2}{||\coeffSHVec||_2},
\end{equation}
where the ground truth SH coefficients $\coeffSHVec$ are calculated from the GMM without noise added, and the NRMSE reconstruction error at the diffusion signal sample locations given by,
\begin{equation}
\label{eq:spatial_error}
\rm{NRMSE}_{\mathbf{d}}  = \frac{||\hat{\diffSigVector}-\diffSigVector ||_2}{||\diffSigVector||_2}.
\end{equation}
Here $\rm{NRMSE}_{\coeffSHVec}$ and $\rm{NRMSE}_{\mathbf{d}} $ are calculated for a $b$-value of 4000$\rm{s/mm}^2$. Using~\figref{fig:SH_bandlimit} it can be determined that the maximum SH degree $\ell$ needed at this $b$-value is $L=8$. 

\begin{figure*}[!h]
  \centering
  \subfloat[SNR=10, CA = $30^\circ$]{ 
  \includegraphics[width=0.33\textwidth]{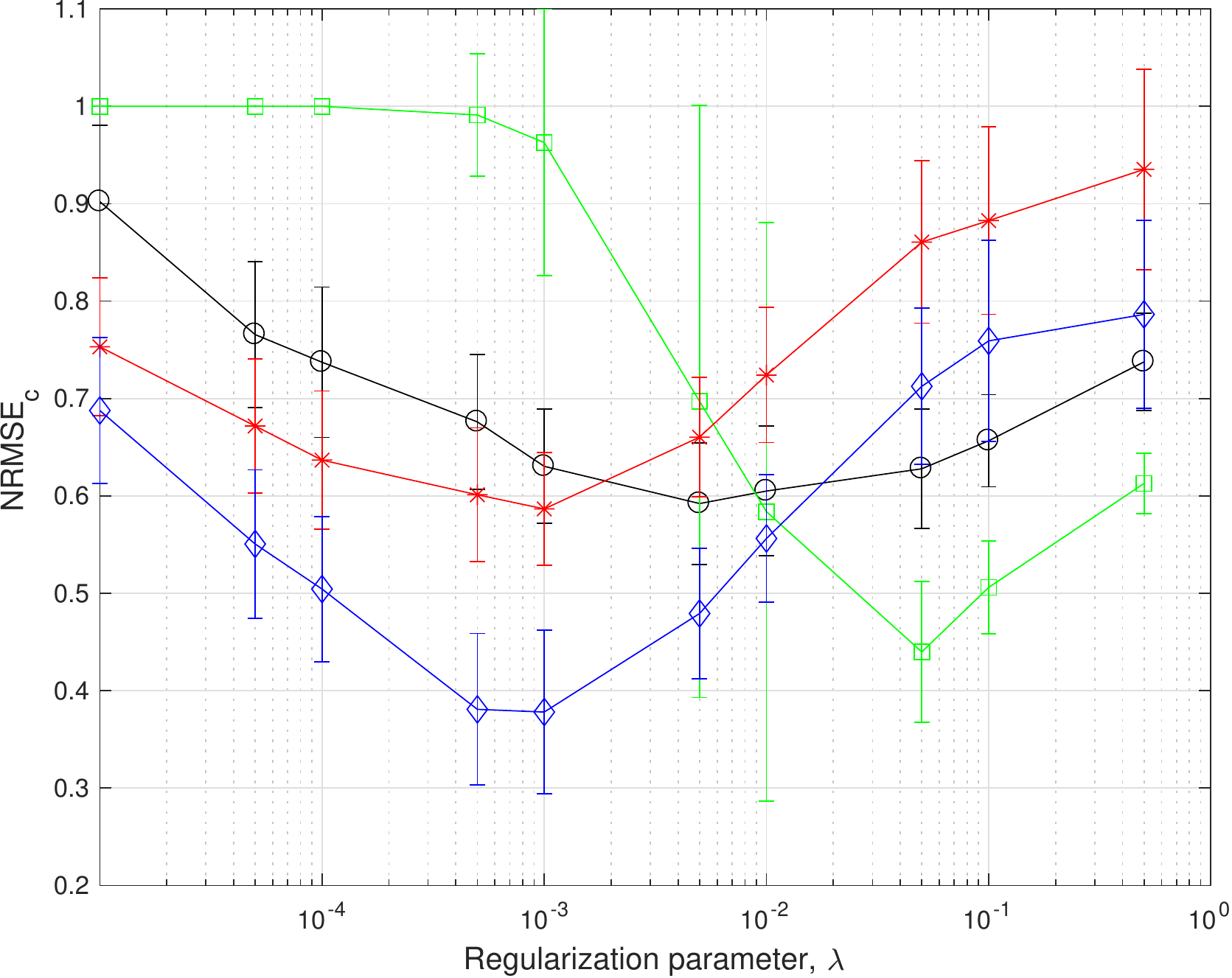}}
  \subfloat[SNR=20, CA = $30^\circ$]{
  \includegraphics[width=0.33\textwidth]{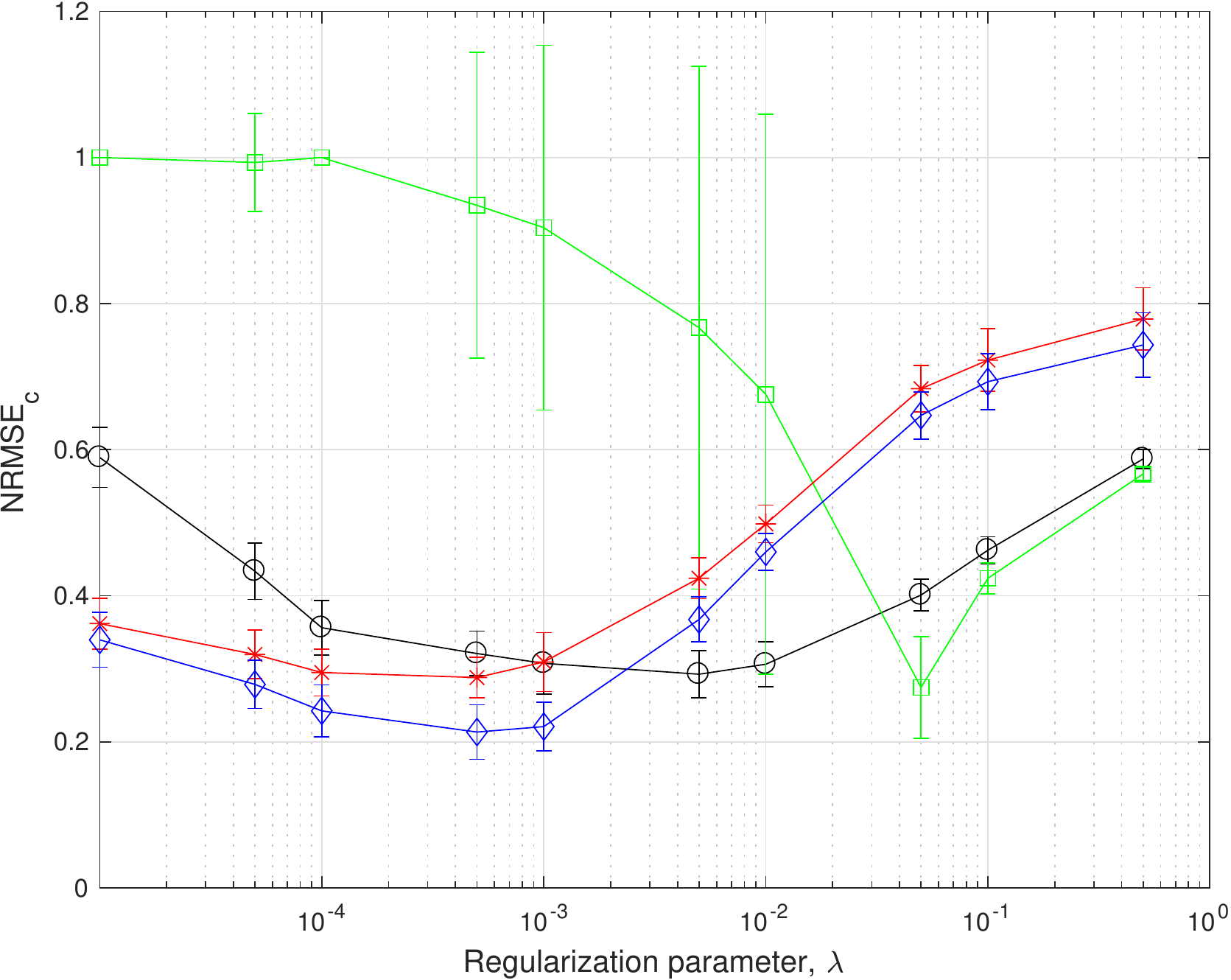}} 
  \subfloat[SNR=30, CA = $30^\circ$]{
  \includegraphics[width=0.33\textwidth]{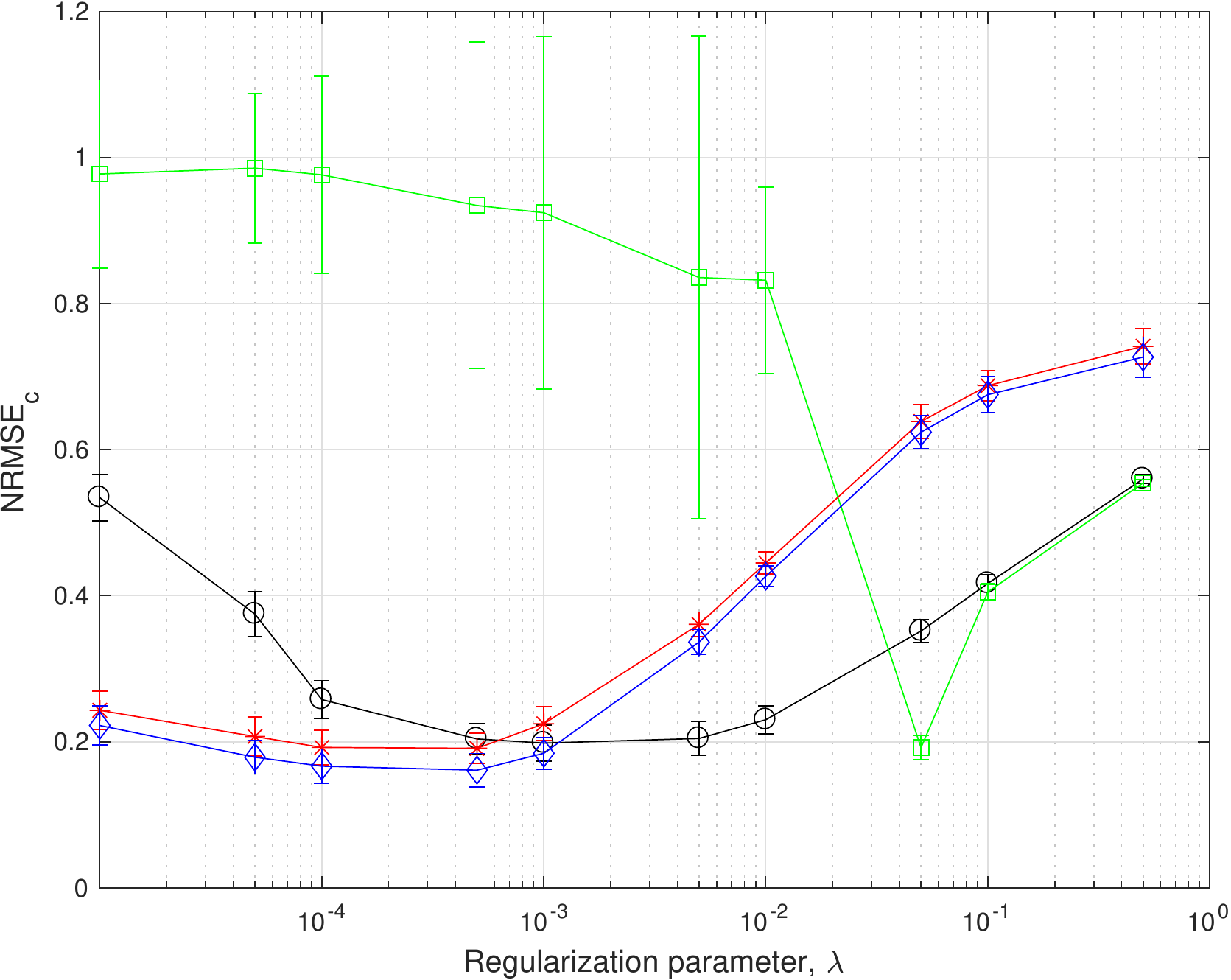}}

  \subfloat[SNR=10, CA = $90^\circ$]{ 
  \includegraphics[width=0.33\textwidth]{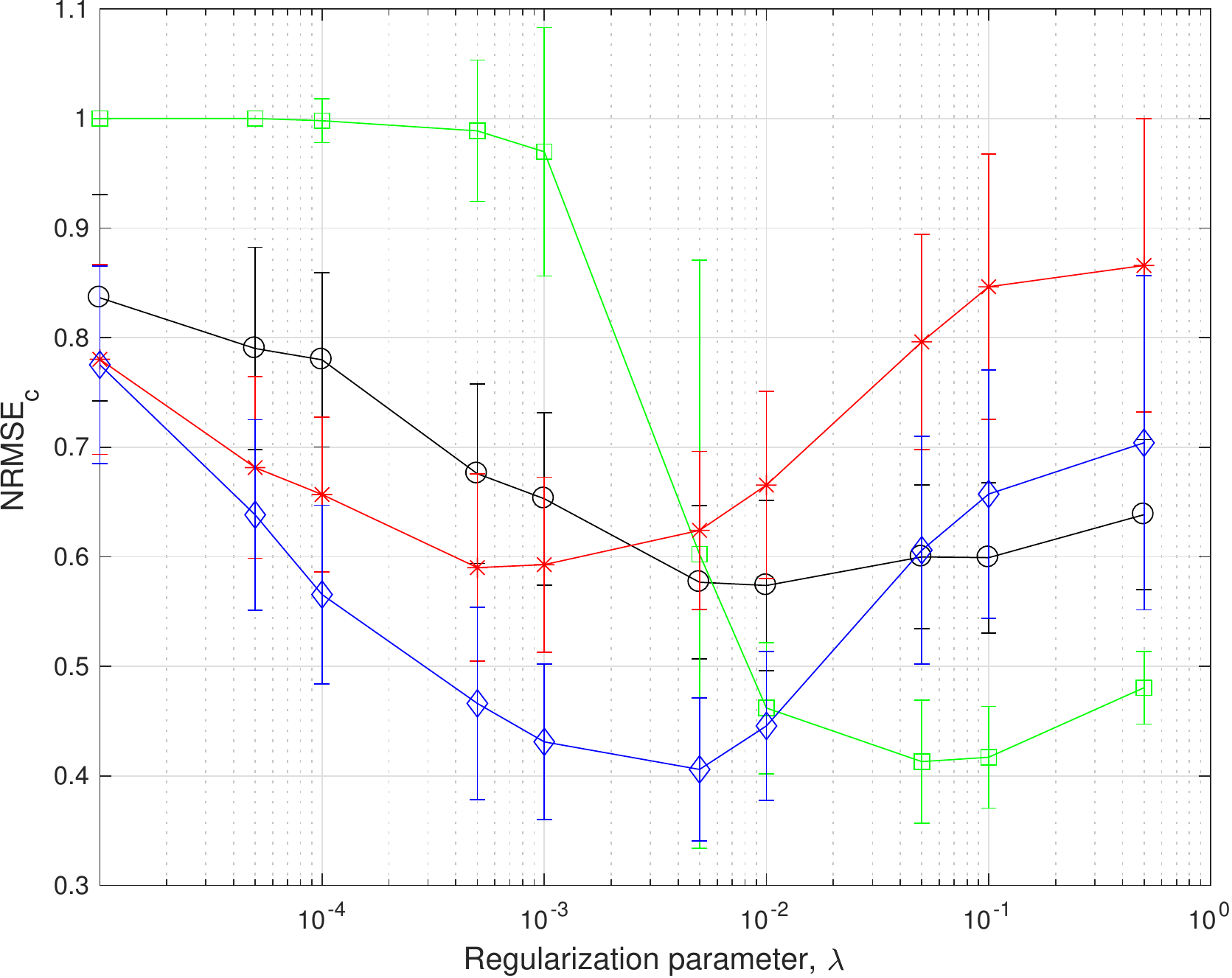}} 
    \subfloat[SNR=20, CA = $90^\circ$]{
  \includegraphics[width=0.33\textwidth]{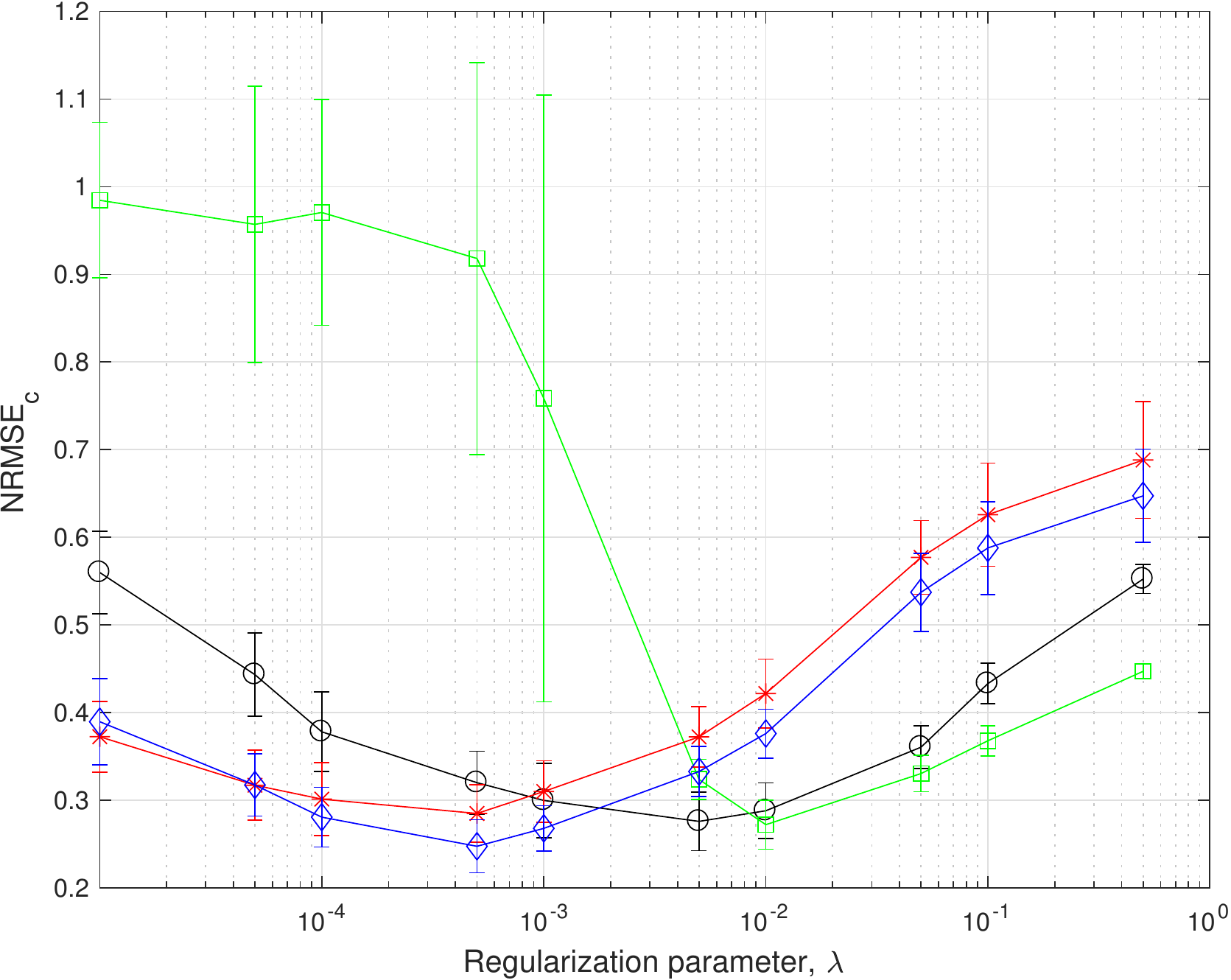}} 
  \subfloat[SNR=30, CA = $90^\circ$]{
  \includegraphics[width=0.33\textwidth]{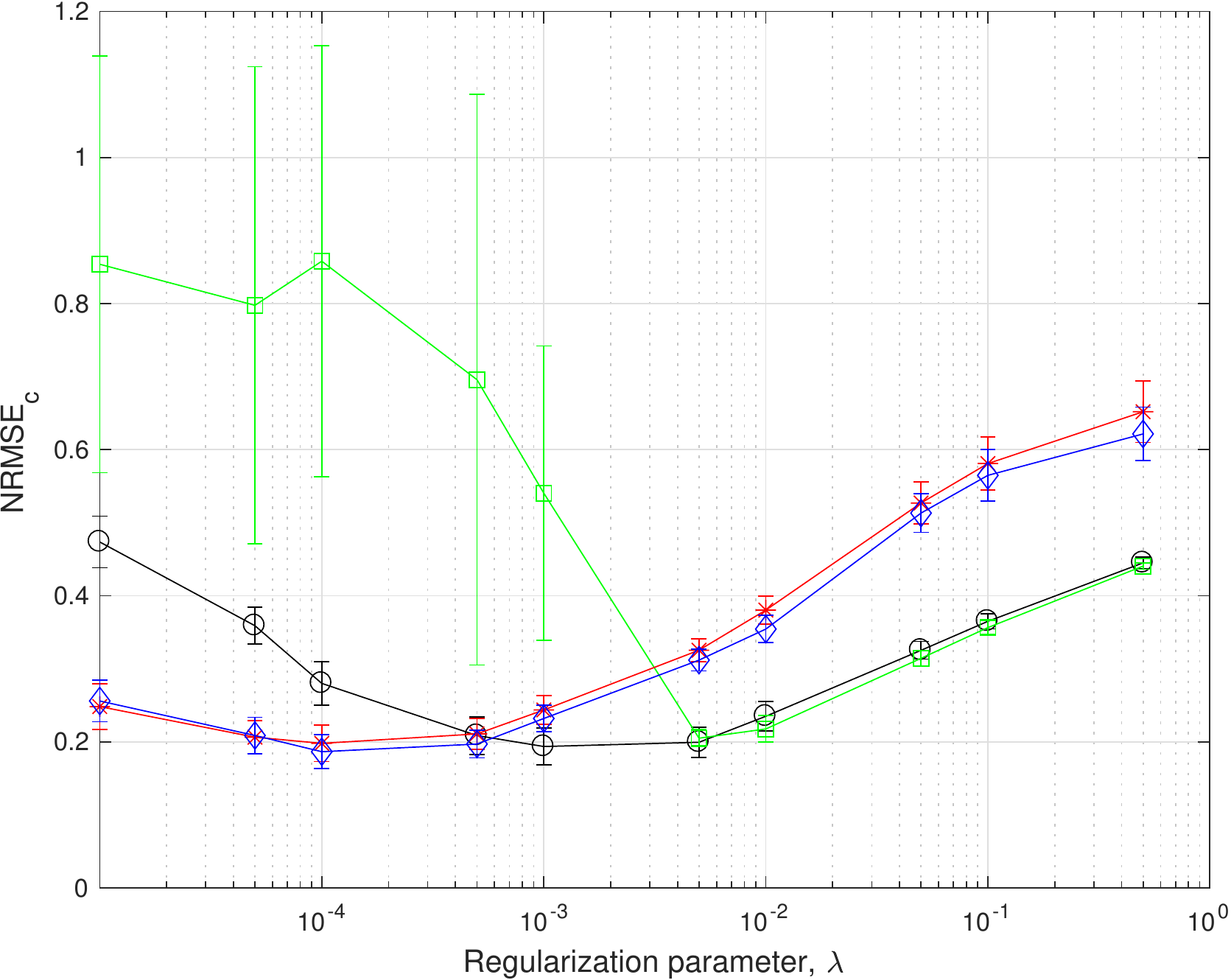}} \hfil
  
  \subfloat{\includegraphics[width=0.17\textwidth]{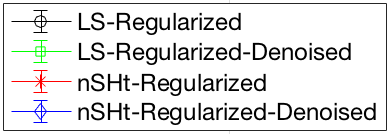}}
    \caption{{\bf Single-shell sampling scheme reconstruction accuracy on GMM simulation with crossing fibers.} Normalized root mean-squared error of the SH coefficients, $\rm{NRMSE}_{\coeffSHVec}$, obtained using LS-Regularized, LS-Regularized-Denoised, nSHt-Regularized and nSHt-Regularized-Denoised for Gaussian mixture model (GMM) simulations with two fiber compartments with crossing angles (CA) $30^\circ$ (first row) and $90^\circ$ (second row) with SNR 10, 20, and 30. }
  \label{fig:NRMSE_ss_CA}
\end{figure*}

\begin{figure*}[!h]
  \centering
  \subfloat[SNR=10, FA = 0.6]{ 
  \includegraphics[width=0.33\textwidth]{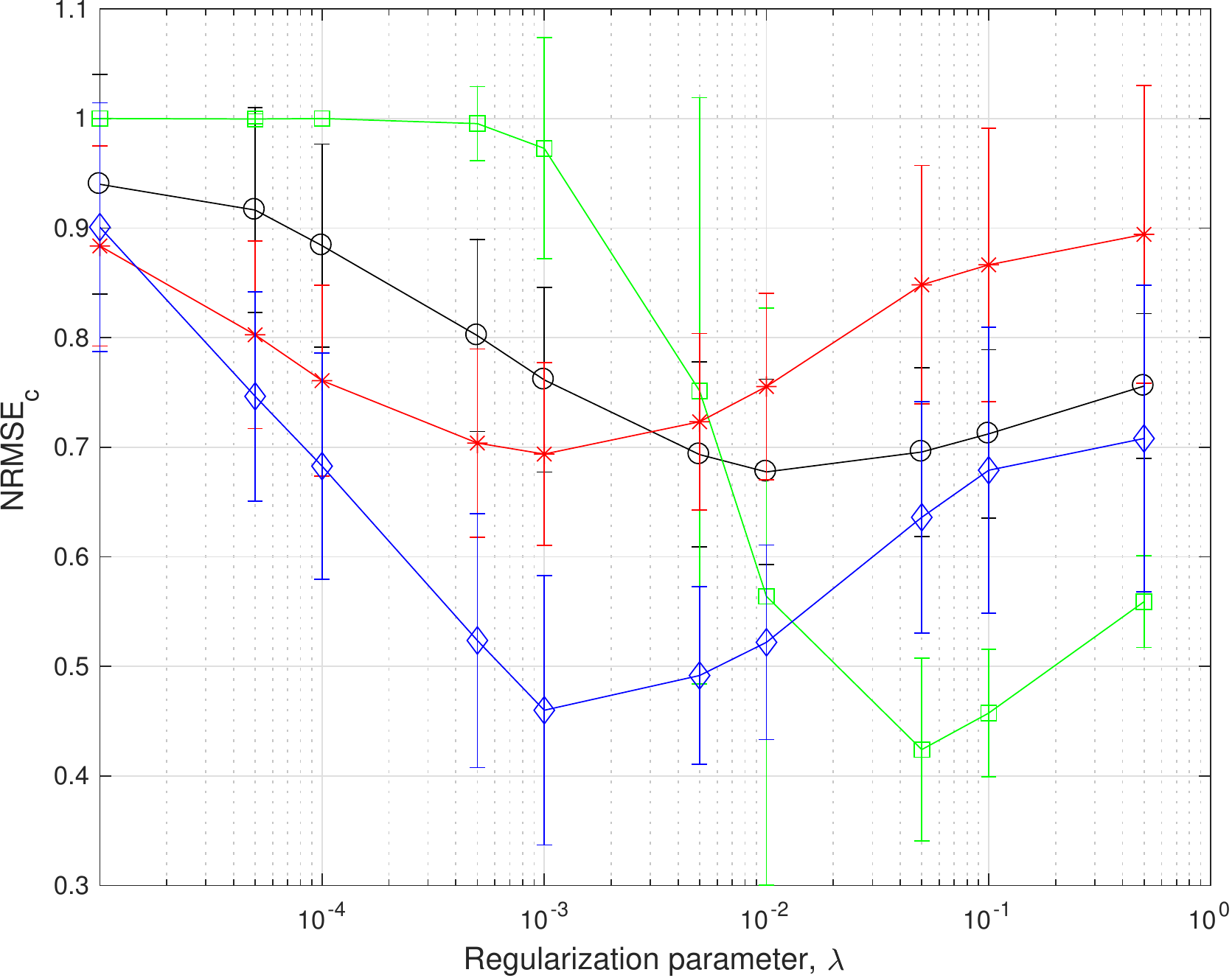}}
  \subfloat[SNR=20, FA = 0.6]{
  \includegraphics[width=0.33\textwidth]{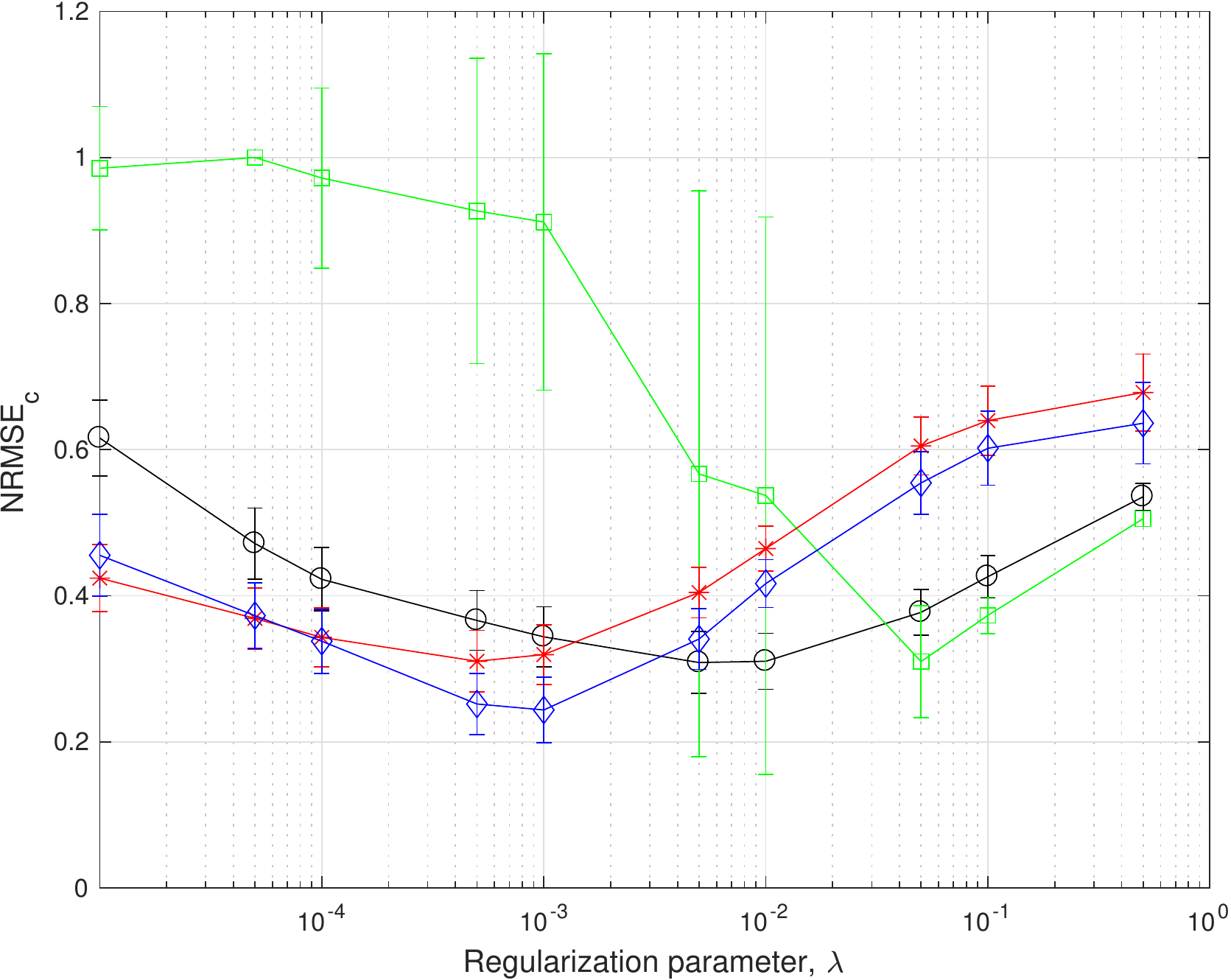}} 
  \subfloat[SNR=30, FA = 0.6]{
  \includegraphics[width=0.33\textwidth]{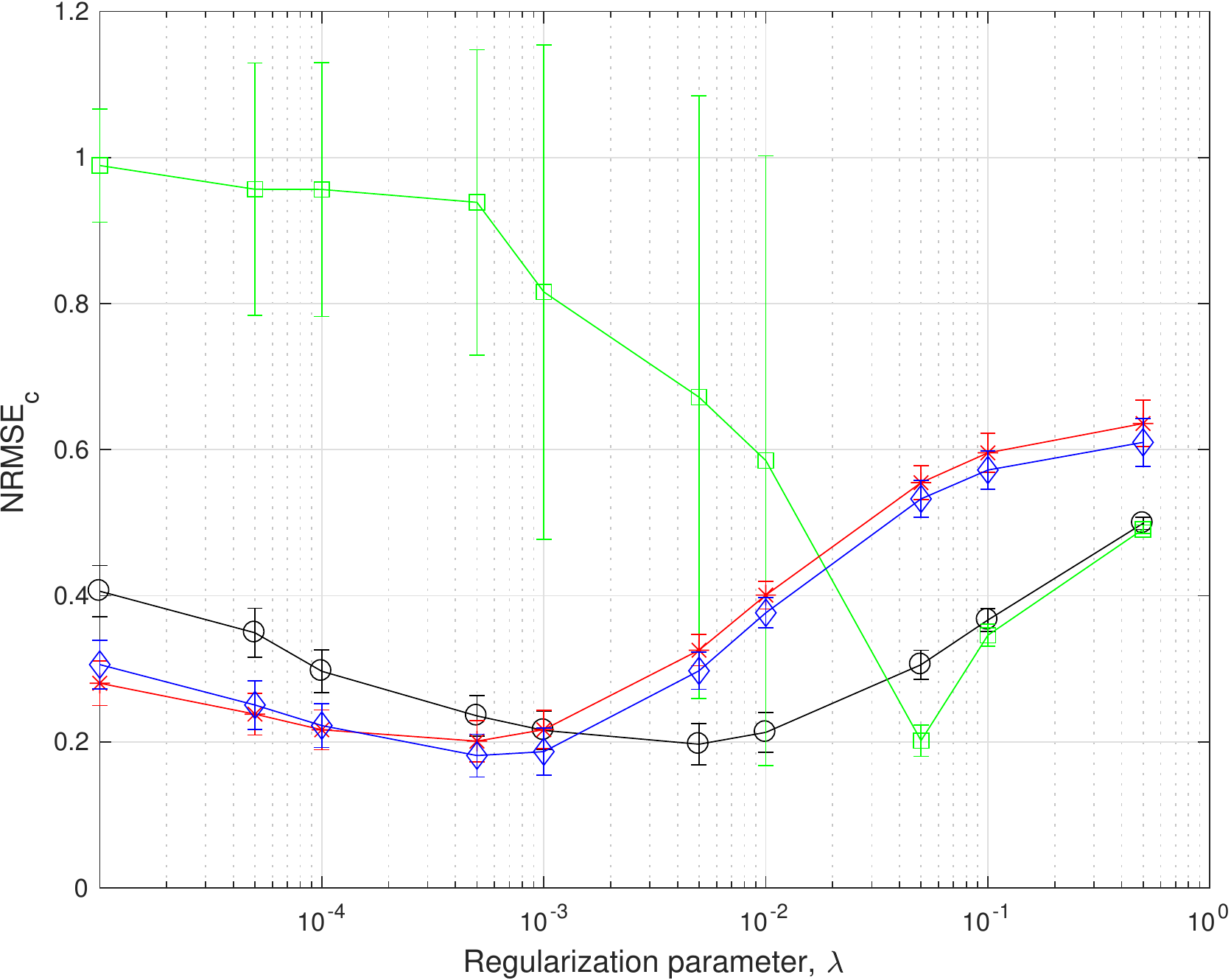}}
  
  \subfloat[SNR=10, FA = 0.8]{ 
  \includegraphics[width=0.33\textwidth]{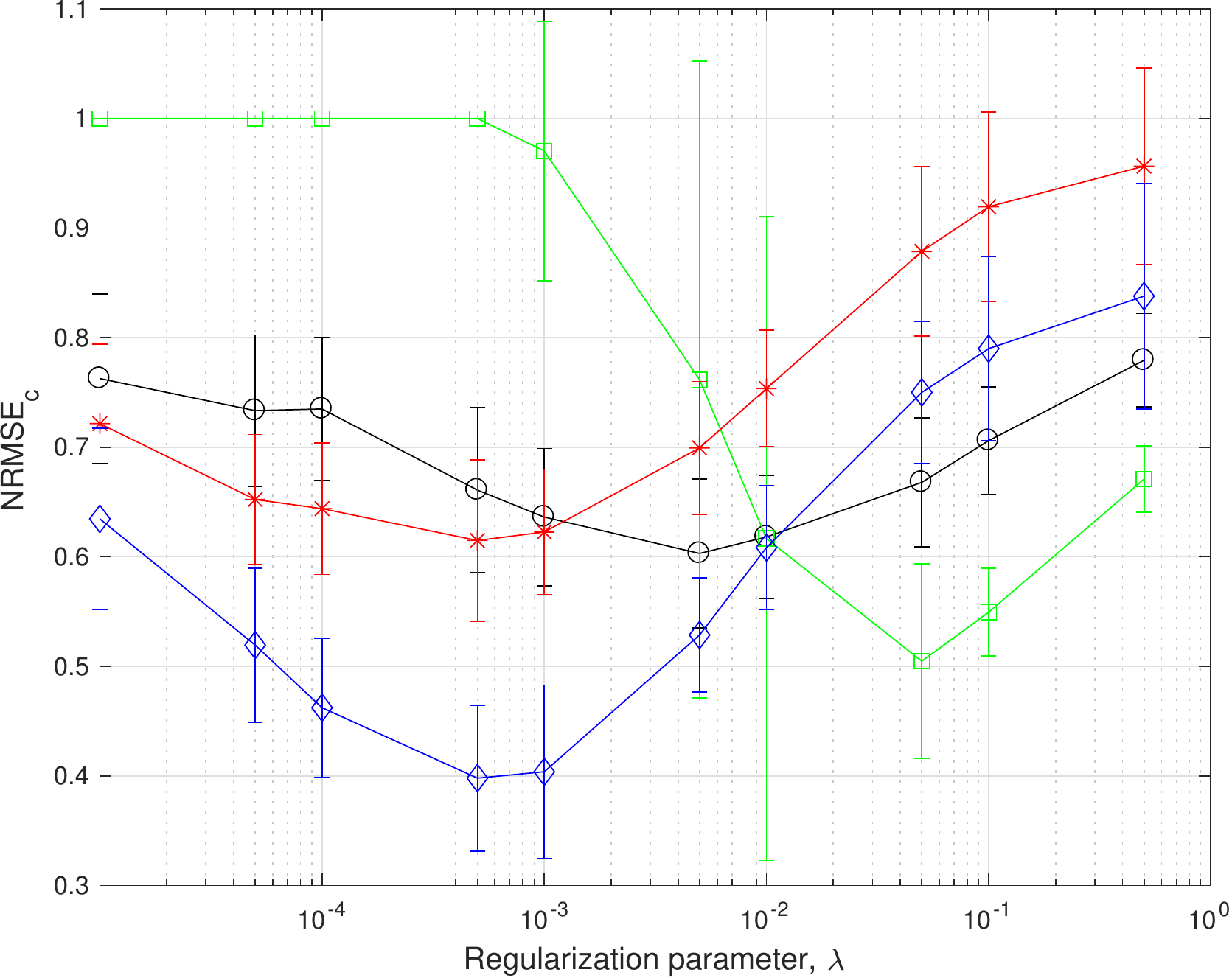}} 
    \subfloat[SNR=20, FA = 0.8]{
  \includegraphics[width=0.33\textwidth]{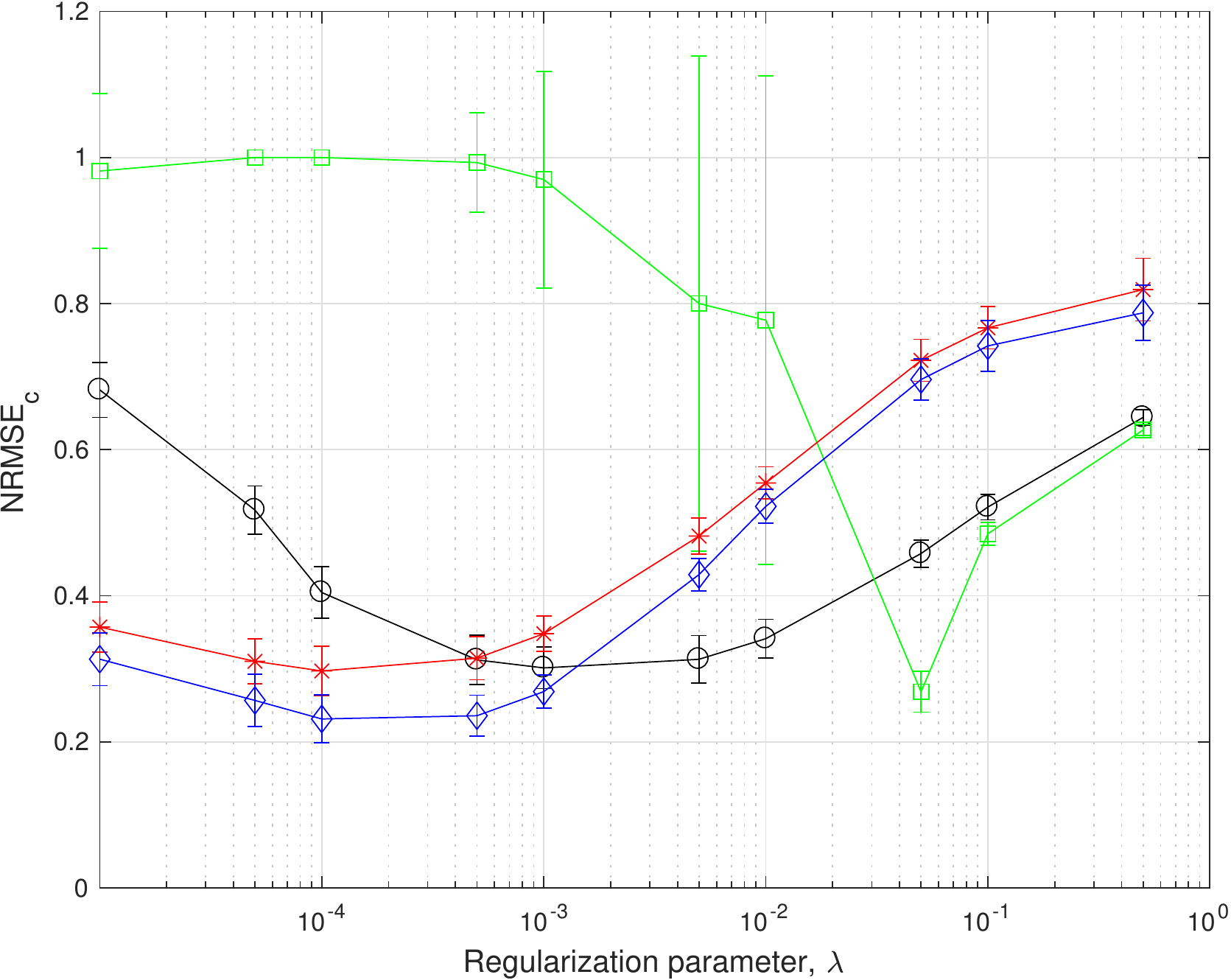}} 
  \subfloat[SNR=30, FA = 0.8]{
  \includegraphics[width=0.33\textwidth]{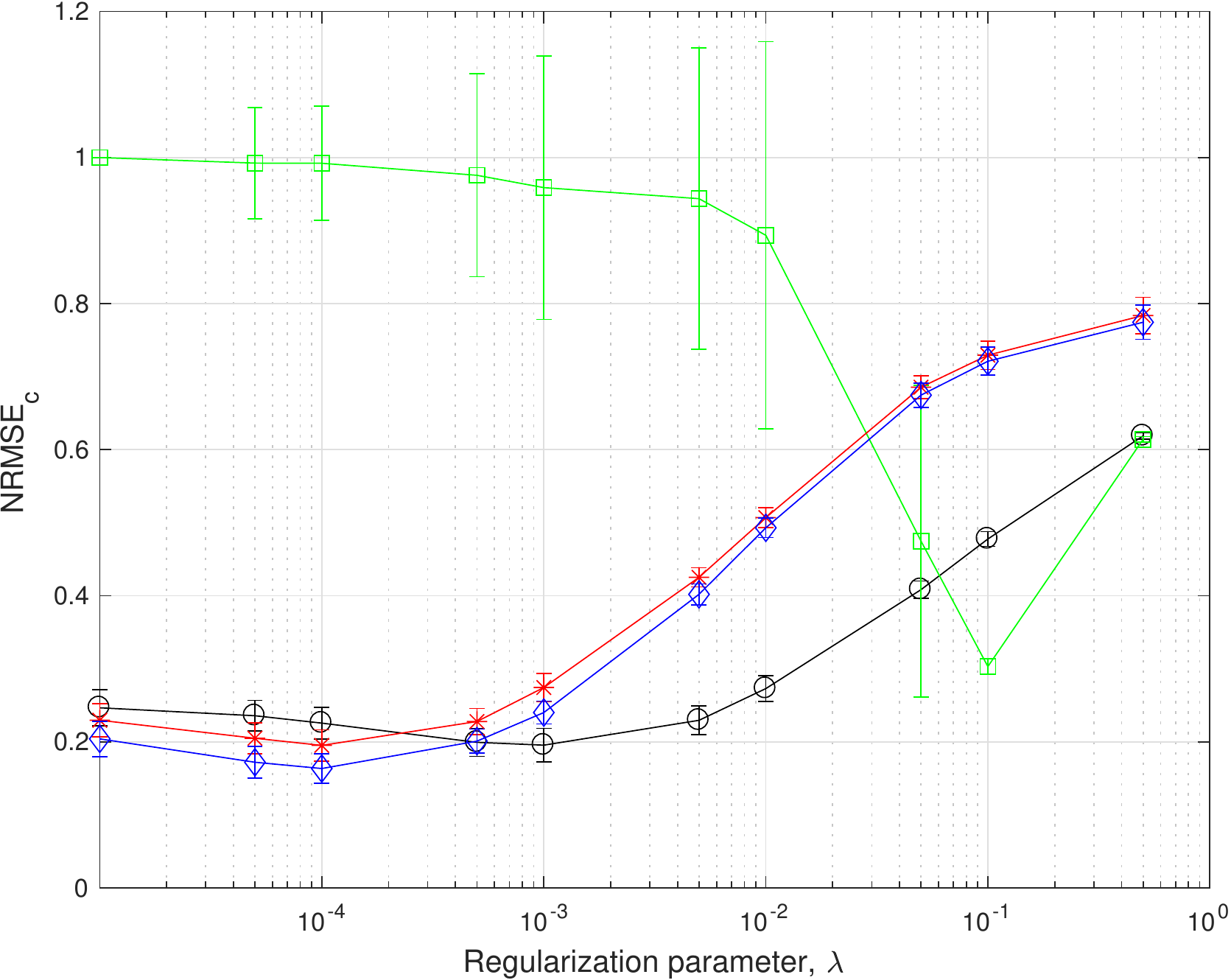}} \hfil 
  \subfloat{\includegraphics[width=0.17\textwidth]{legend_ss}}
   \caption{{\bf Single-shell sampling scheme reconstruction accuracy on GMM simulation with a single fiber.} Normalized root mean-squared error of the SH coefficients, $\rm{NRMSE}_{\coeffSHVec}$, obtained using LS-Regularized, LS-Regularized-Denoised, nSHt-Regularized and nSHt-Regularized-Denoised for GMM simulations with a single fiber compartment with fractional anisotropies (FA) $0.6$ (first row) and $0.8$ (second row) with SNR 10, 20, and 30. }
  \label{fig:NRMSE_ss_FA}
\end{figure*}

$\rm{NRMSE}_{\coeffSHVec}$ as a function of regularization parameter $\lambda$ for the first synthetic data-set where the crossing angle is changed is shown in \figref{fig:NRMSE_ss_CA} and for the second data-set where the FA is changed is shown in \figref{fig:NRMSE_ss_FA}. Due to space constraints, only results for crossing-angles of $30^\circ$ and $90^\circ$, and for FAs of 0.6 and 0.8 are shown here. Results for the other crossing angles and FAs are contained in \nameref{S1_File}. As $\rm{NRMSE}_{\coeffSHVec}$ and $\rm{NRMSE}_{\mathbf{d}}$ have the same trend, only $\rm{NRMSE}_{\coeffSHVec}$ is included in the paper, the figures for $\rm{NRMSE}_{\mathbf{d}}$ are also contained in \nameref{S1_File}. 

\subsubsection*{Multi-shell}\label{Sec:val_ms}
Here we evaluate the proposed multi-shell scheme with regularized novel SPFt presented in Proposed Multi-shell Sampling Scheme - Regularization Subsection, denoted nSPFt-Reg, and also the scheme additionally modeling the non-Gaussian noise presented in Proposed Multi-shell Sampling Scheme - Non-Gaussian Noise Removal  Subsection, denoted nSPFt-Reg-Denoised.

We sample the GMM of the diffusion signal using the multi-shell sampling grid.  Here, we use a maximum $b$-value of 4000 $\rm{s/mm}^2$ and $N=3$ (4 shells), as it is found in \cite{Assemlal:2009b} that this number of shells is sufficient for convergence to the ground truth when the signal is Gaussian or bi-Gaussian, resulting in shells at $b = 206, 847, 2018$ and $4000$ $\rm{s/mm}^2$. It should be noted that the scheme can be designed for any maximum $b$-value and any number of shells. The SH band-limit is determined using \figref{fig:SH_bandlimit} to be $L(s) =[ 2, 4, 6, 8]$ for the inner most (smaller $b$-value) to outer most shell, giving a total of 94 samples, calculated using Eq~\eqref{Eq:number_samples_antipodal}.

As an evaluation metric, we use the NRMSE of the estimated SPF coefficients,
\begin{equation}
\label{Eq:SPF_error}
\rm{NRMSE}_{\coeffSPFVec} = \frac{||\hat{\coeffSPFVec}-\coeffSPFVec ||_2}{||\coeffSPFVec||_2},
\end{equation}
where the ground truth SPF coefficients $\coeffSPFVec$ are calculated from the GMM without noise added,
and the NRMSE reconstruction error at the diffusion signal sample locations is given by Eq~\eqref{eq:spatial_error}.

\begin{figure*}[!h]
\centering
  \subfloat[SNR=10]{ 
  \includegraphics[trim=30 10 45 25,clip,width=0.33\textwidth]{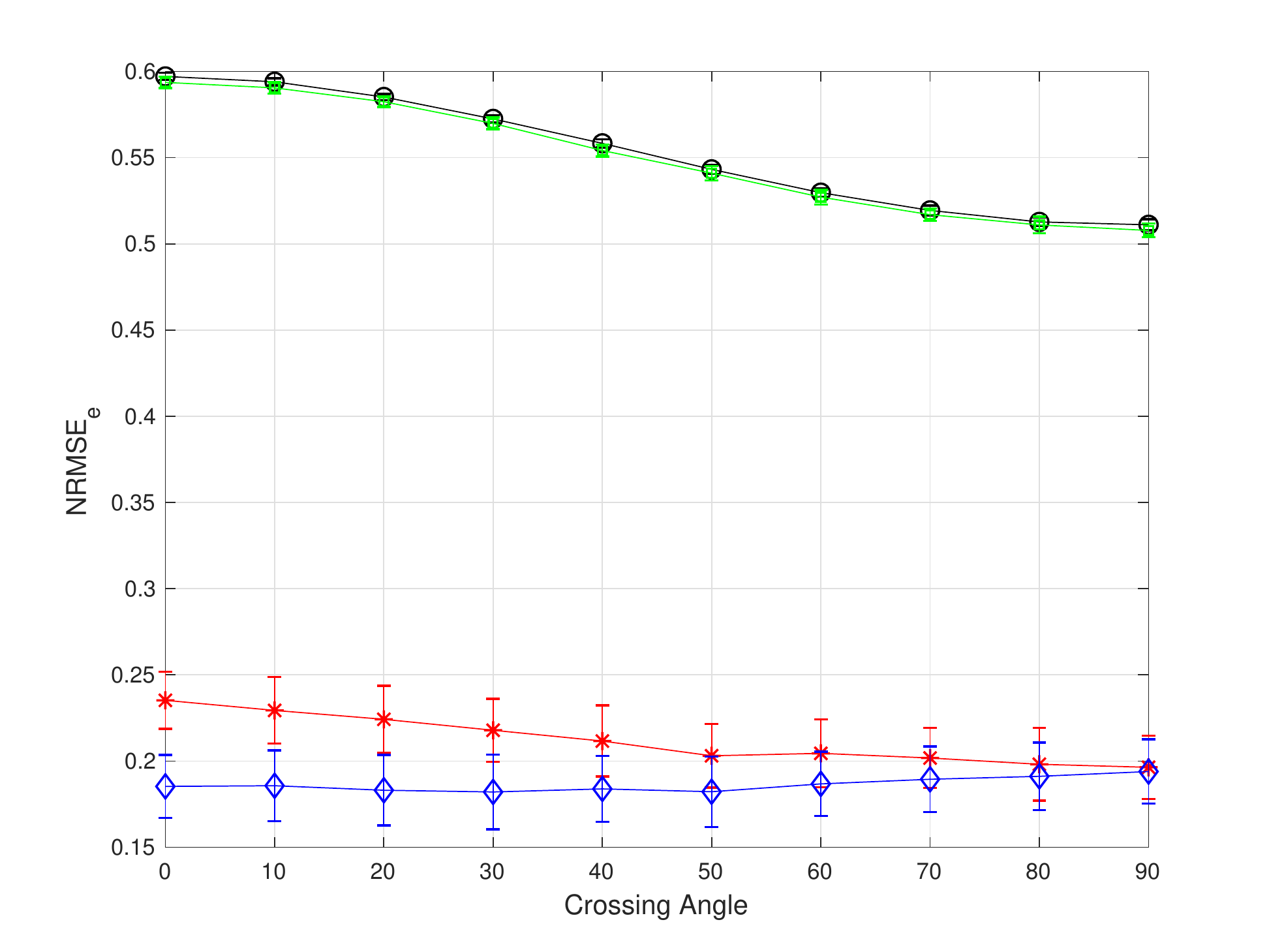}}
  \subfloat[SNR=20]{
  \includegraphics[trim=30 10 45 25,clip,width=0.33\textwidth]{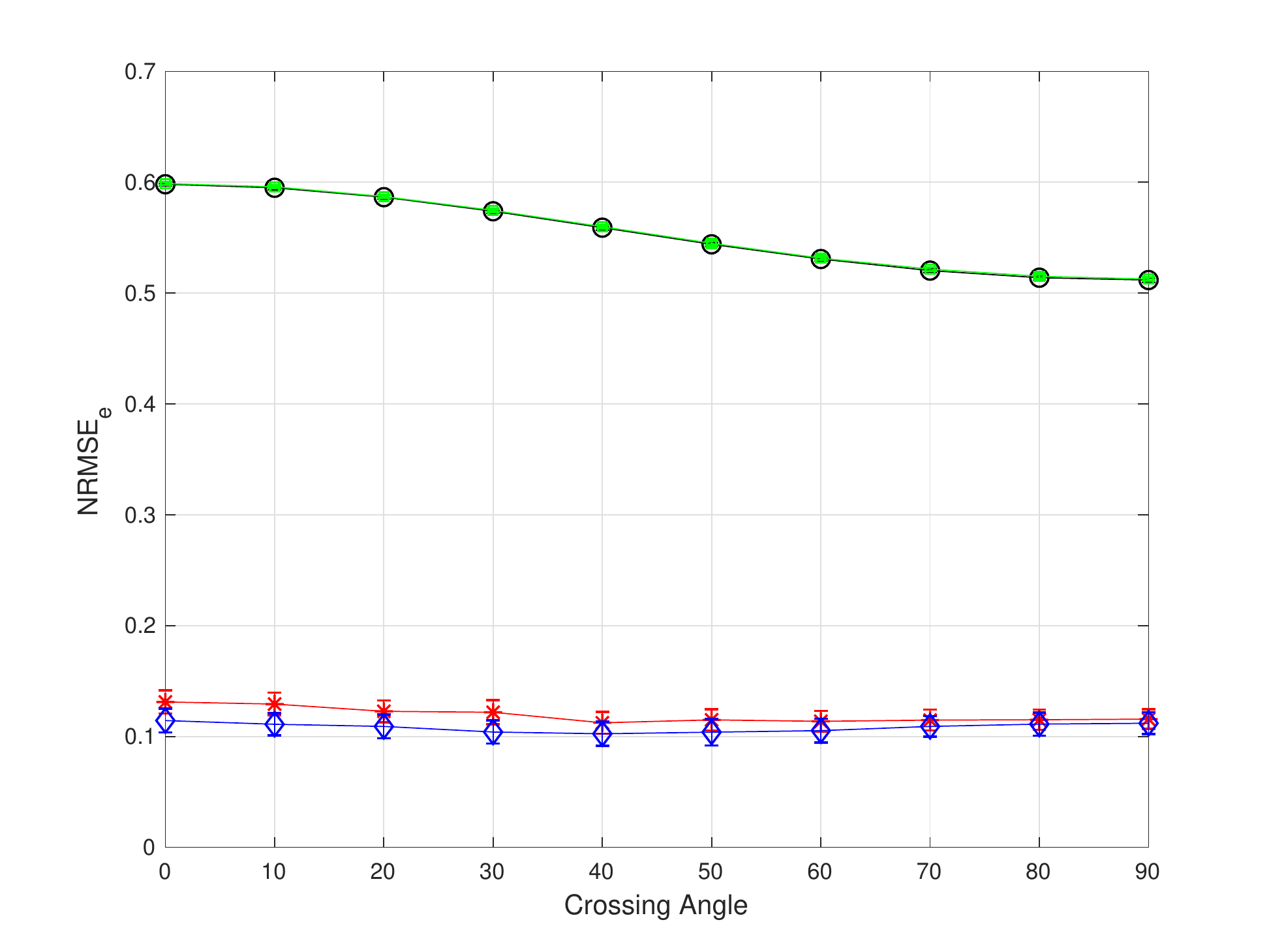}} 
  \subfloat[SNR=30]{
  \includegraphics[trim=30 10 45 25,clip,width=0.33\textwidth]{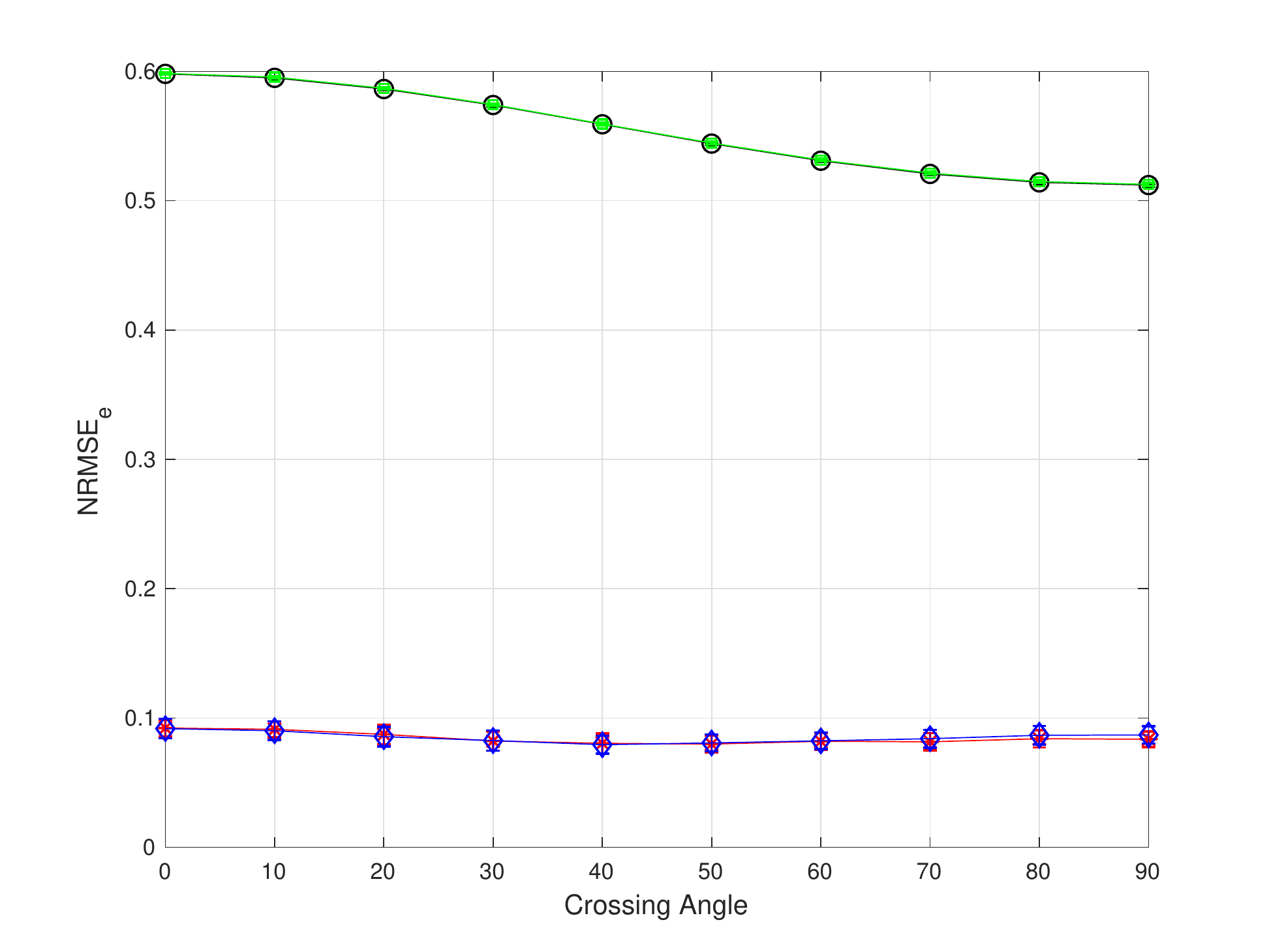}}
  
  \subfloat[SNR=10]{ 
  \includegraphics[trim=30 10 45 25,clip,width=0.33\textwidth]{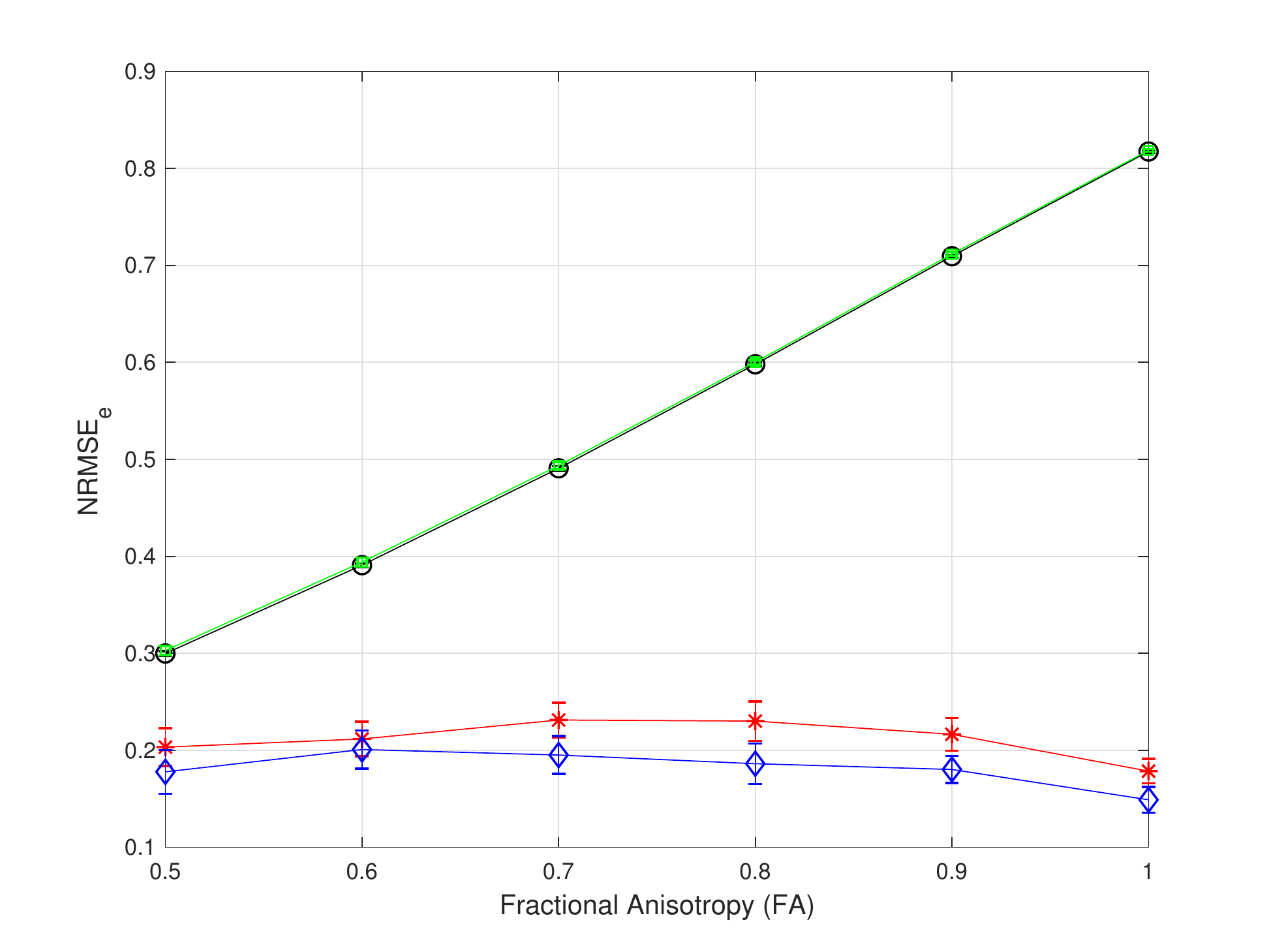}} 
    \subfloat[SNR=20]{
  \includegraphics[trim=30 10 45 25,clip,width=0.33\textwidth]{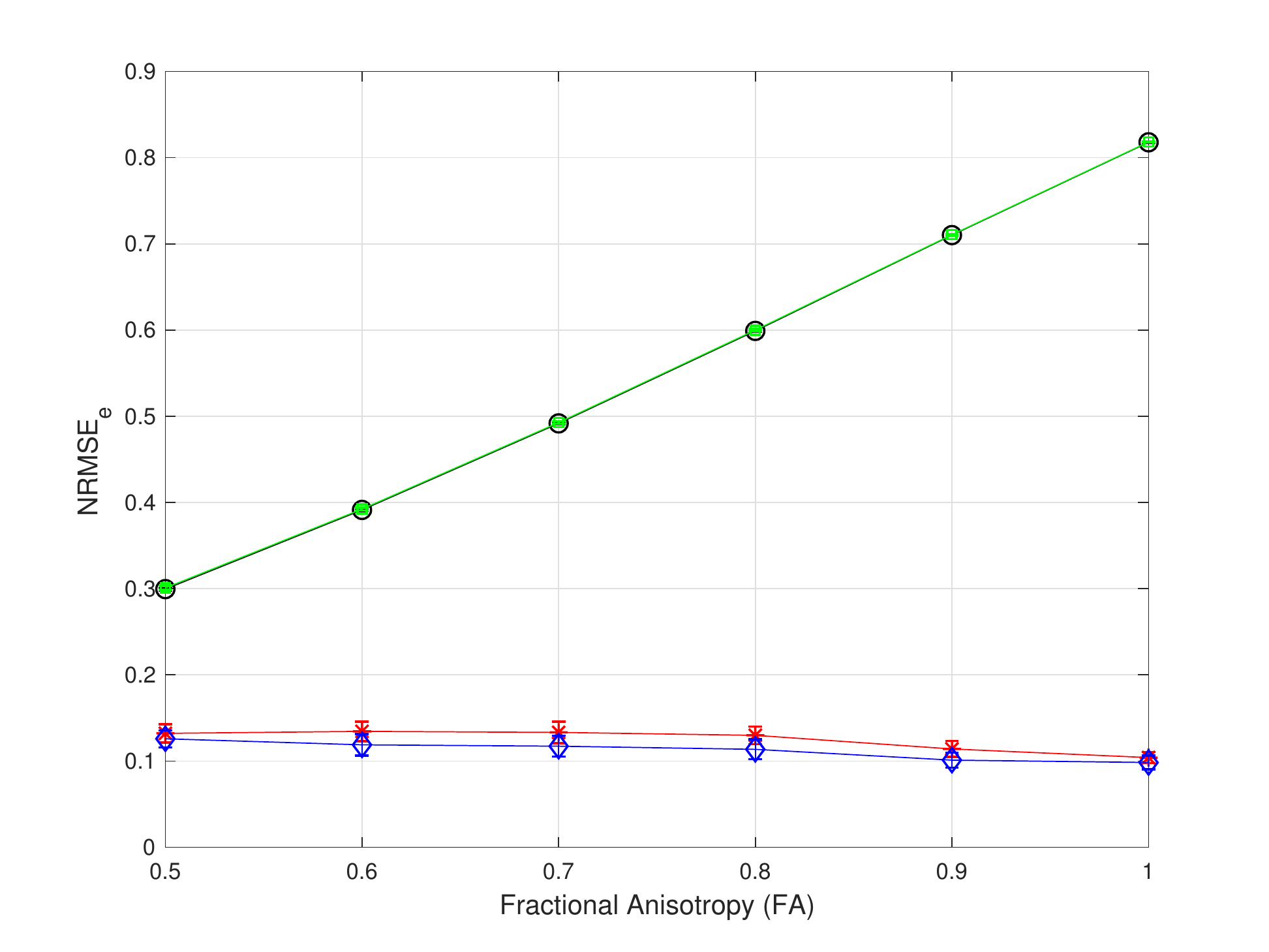}} 
  \subfloat[SNR=30]{
  \includegraphics[trim=30 10 45 25,clip,width=0.33\textwidth]{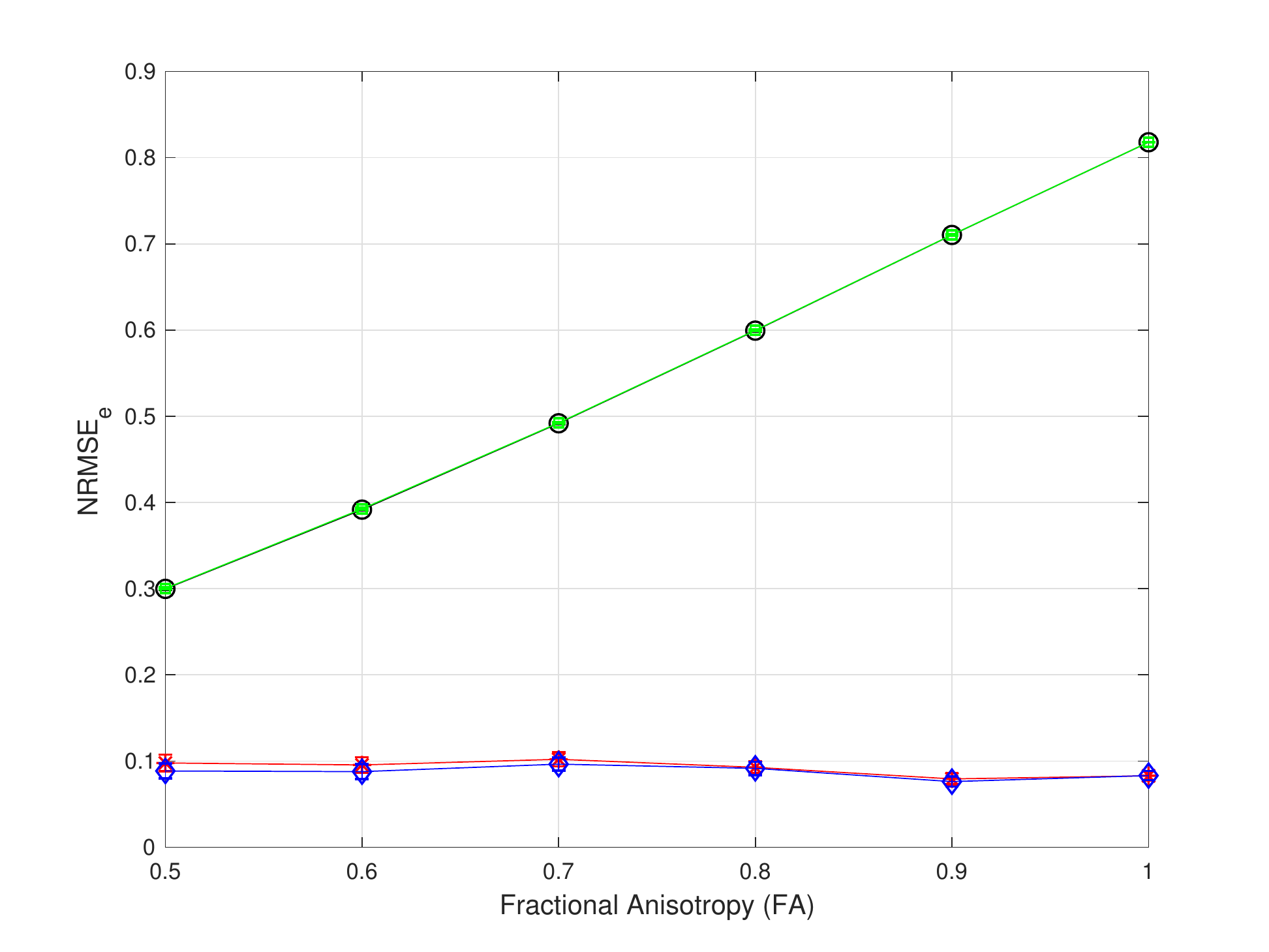}}  \hfil
  \subfloat{\includegraphics[width=0.17\textwidth]{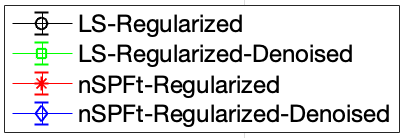}}
  \caption{{\bf Multi-shell sampling scheme reconstruction accuracy on GMM simulation.} Normalized root mean-squared error of the SPF coefficients, $\rm{NRMSE}_{\coeffSPFVec}$, obtained using LS-Regularized, LS-Regularized-Denoised, nSPFt-Regularized and nSPFt-Regularized-Denoised. In the first row $\rm{NRMSE}_{\coeffSPFVec}$ is shown  for GMM simulations with two fiber compartments with crossing angles $0^\circ$ to $90^\circ$ with SNR 10 (a), 20 (b), and 30 (c). In the second row $\rm{NRMSE}_{\coeffSPFVec}$ is also shown for GMM with a single fiber compartment with FA of 0.5 to 1 with SNR 10 (d), 20 (e), and 30 (f).}
  \label{fig:NRMSE_spectral_error}
\end{figure*}

$\rm{NRMSE}_{\coeffSPFVec}$ for the synthetic data-set where the crossing angle was changed is shown in \figref{fig:NRMSE_spectral_error}(a)-(c) and in \figref{fig:NRMSE_spectral_error}(d)-(f) for the data-set where the FA was changed. As $\rm{NRMSE}_{\coeffSPFVec}$ and $\rm{NRMSE}_{\mathbf{d}}$ have the same trend, only $\rm{NRMSE}_{\coeffSPFVec}$ is included in the paper, the figures for $\rm{NRMSE}_{\mathbf{d}}$ are contained in \nameref{S1_File}. The regularization parameters $\lambda_\ell$ and $\lambda_n$ were chosen individually for each of the 4 methods compared in \figref{fig:NRMSE_spectral_error} to minimize the NRMSE.

\subsection*{Real data validation}\label{Sec:real_data}
Single-shell data has been acquired from a young healthy adult volunteer using a Siemens Prisma working at 3T with a head coil with 64 channels with adaptive combine method, resulting in the noise having a Rician distribution~\cite{Aja:2016,Varadarajan:2015}. 45 measurements were collected using the single-shell sampling grid with $L=8$ with a $b$-value of 4000$\rm{s/mm}^2$ and a spatial resolution of $1.8 \times 1.8 \times 1.8$ mm.

\begin{figure*}[!h]
  \centering
   \subfloat[]{
    \includegraphics[width=0.18\textwidth]{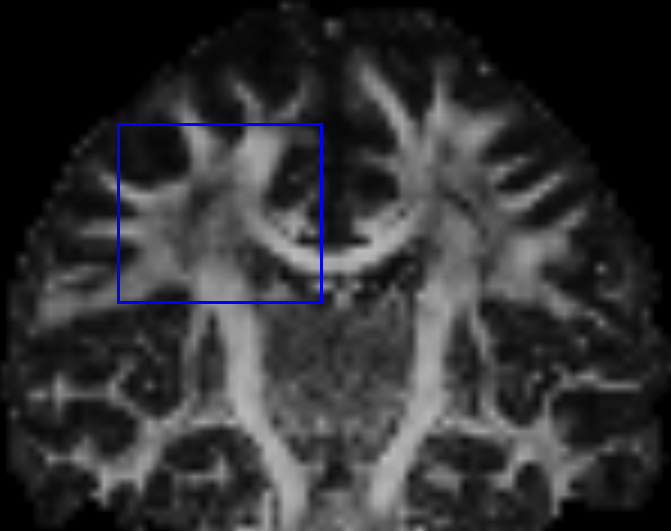}
  }\hfil
\subfloat[]{  
  \includegraphics[trim=220 120 180 105,clip,width=0.5\textwidth]{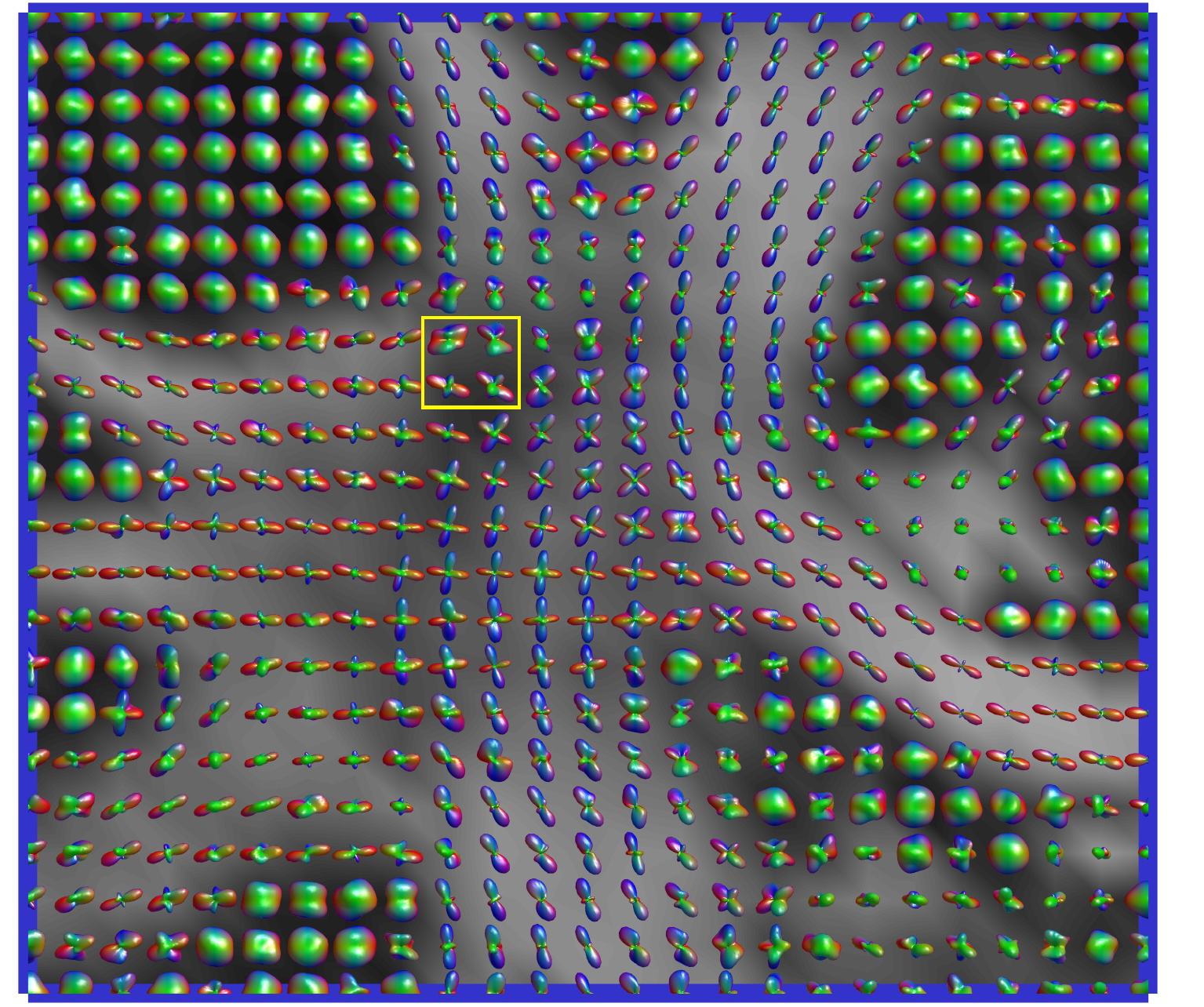} 
  }
\caption{{\bf ODF profiles estimated from real data using proposed sampling scheme.} (a) The FA of a coronal slice with the blue box surrounding the a region of interest in the left hemisphere of the brain. (b) Visualization of the ODFs estimated using the proposed method nSHt-Regularized in the region of interest.}
\label{fig:odfs_proposed}
\end{figure*}

\begin{figure*}[!h]
   \centering
      \centering
   \vspace{-4mm}
 \subfloat[]{
  \includegraphics[trim=220 120 180 105,clip,width=0.5\textwidth]{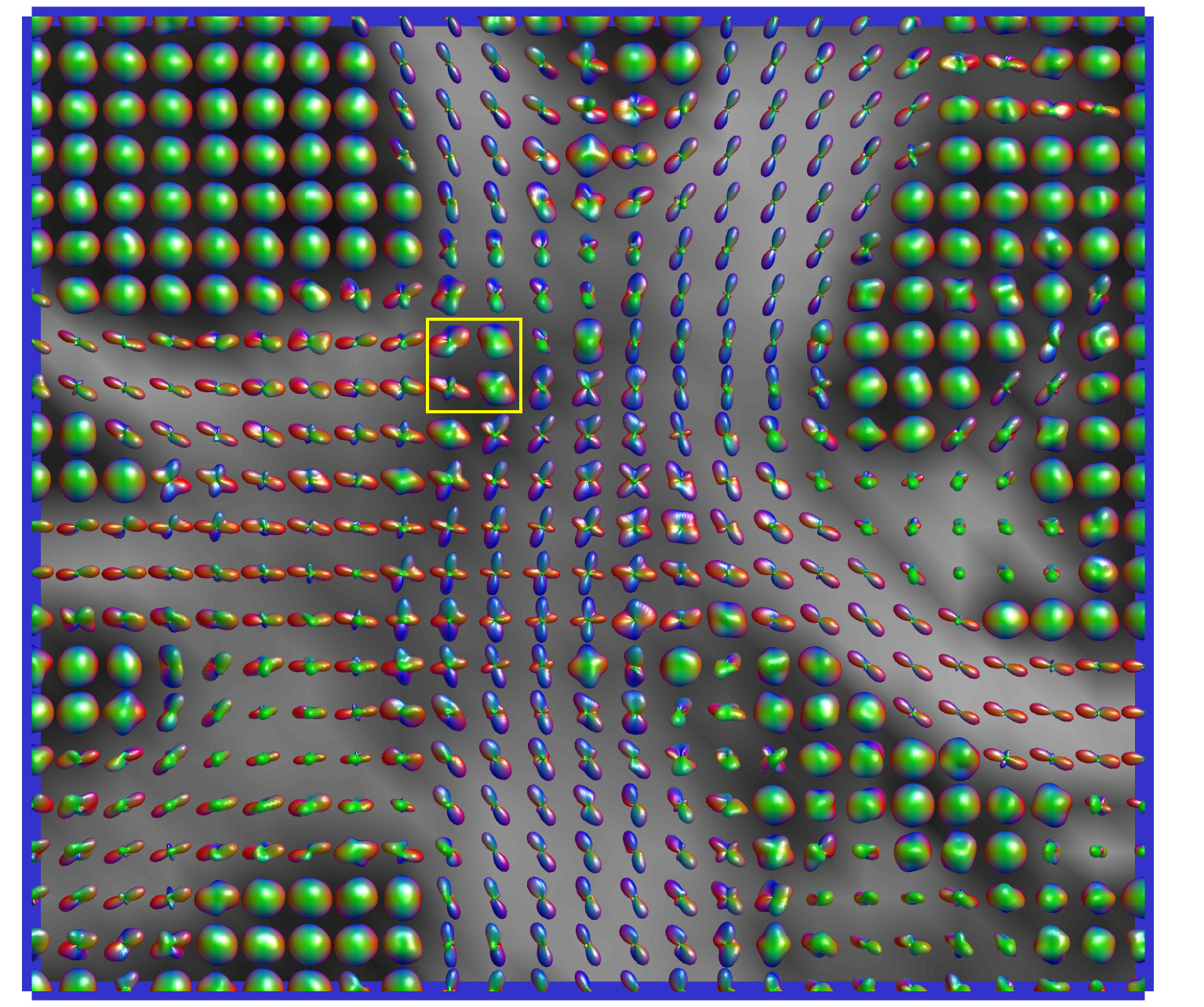}
} 
 \subfloat[]{
   \includegraphics[trim=220 120 180 105,clip,width=0.5\textwidth]{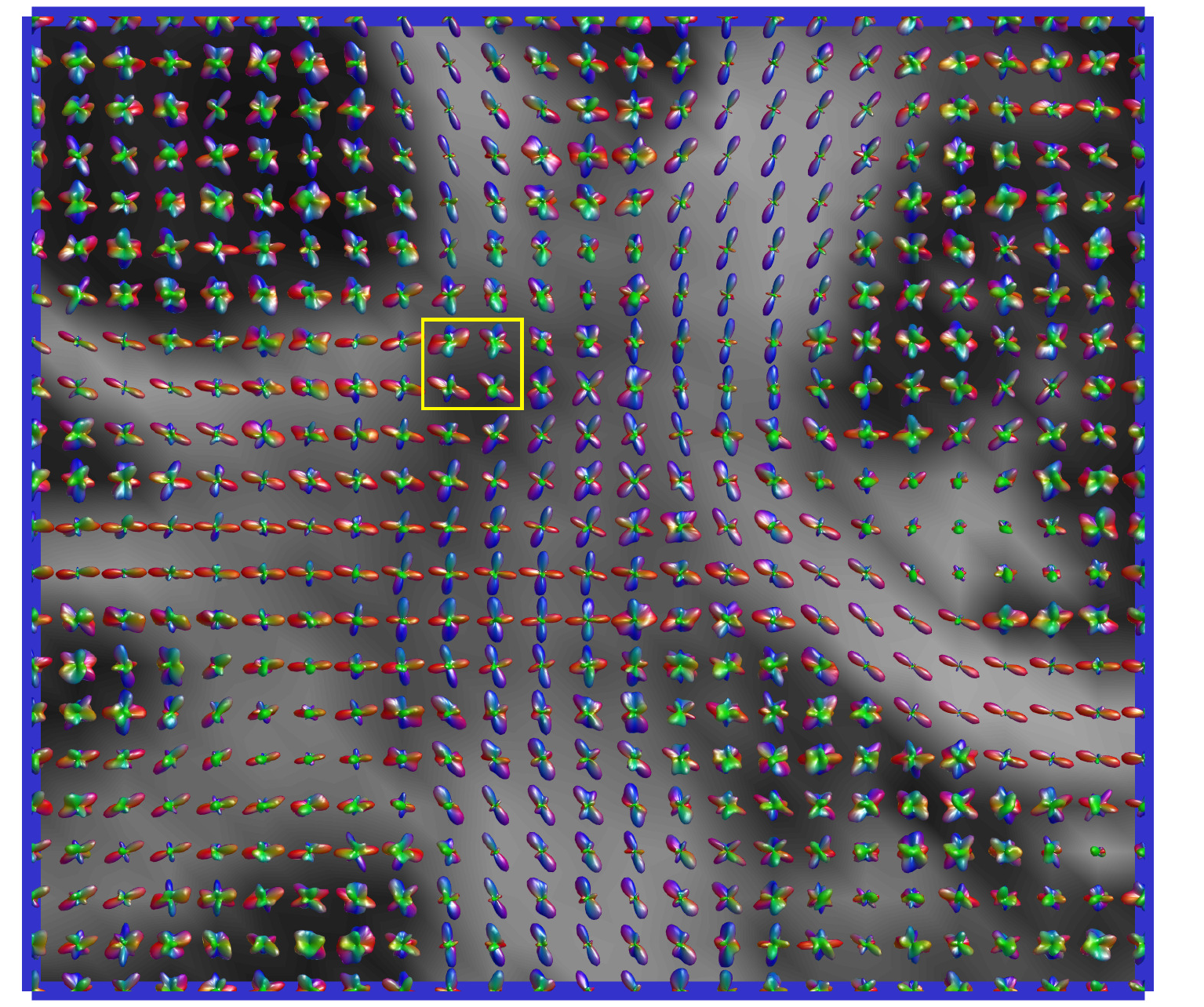}
 }
   \caption{{\bf ODF profiles estimated from real data using least-squares reconstruction. } Visualization of the ODFs estimated using standard regularized least-squares reconstruction LS-Regularized in a region of interest in the left hemisphere of the brain shown in \figref{fig:odfs_proposed}(a). In (a) optimal regularization for least-squares is used and in (b) the same regularization is used as for \figref{fig:odfs_proposed}(b).}
\label{fig:odfs_LS}
\end{figure*}

The proposed sampling schemes reconstruct the diffusion signal, which can then be used to extract its features such as the ODF, which we do here. The SH coefficients of the diffusion signal are computed in every voxel using the single-shell sampling scheme with regularized SHt, nSHt-Regularized. The SH coefficients of the ODF are then calculated from the coefficients of the diffusion signal~\cite{Aganj:2010}. The reconstructed ODFs are shown in~\figref{fig:odfs_proposed}(b) for a coronal region of interest on the left hemisphere of the brain, shown by the blue box in~\figref{fig:odfs_proposed}(a). The ODFs are color coded to show the diffusion directions, with red, blue and green representing diffusion from left to right (along the x-axis), anterior-posterior (y-axis) and inferior-superior (z-axis), respectively. For comparison, the ODFs for the same region obtained using standard regularized least-squares reconstruction method, LS-Regularized, to compute the SH coefficients of the diffusion signal are shown in \figref{fig:odfs_LS}. In  \figref{fig:odfs_LS}(a) the optimal regularization parameter is used for least-squares and in \figref{fig:odfs_LS}(b) the same regularization parameter is used as for the proposed method.  The results obtained using denoising are indistinguishable from just regularizing the solution, due to the relatively high SNR, and are hence not shown. The regularization parameter $\lambda$ is set for each voxel based on what minimizes the NRMSE for the GMM for different FAs, where the FA is approximated by the generalized FA of a voxel.

\section*{Discussion}\label{Sec:discussion}
From the quantitative results shown in \figref{fig:NRMSE_ss_CA}-6 and the qualitative results shown in \figref{fig:odfs_proposed} and 8, it can be seen that the proposed single and multi-shell sampling schemes perform better than the standard regularized least-squares method for reconstructing the diffusion signal when the same number of samples is used.

For single-shell reconstruction, it can be seen in \figref{fig:NRMSE_ss_CA} and 5 that for all FAs, crossing-angles and SNRs, the proposed single-shell scheme with just regularization, nSHt-Regularized, achieves the minimum $\rm{NRMSE}_{\coeffSHVec}$ (and also $\rm{NRMSE}_{\mathbf{d}}$) for a regularization parameter $\lambda$ that is an order of magnitude or two smaller of that required by the least-squares method of reconstruction, with just regularization, LS-Regularized. Likewise, the proposed single-shell scheme with regularization and denoising, nSHt-Regularized-Denoised, achieves the minimum $\rm{NRMSE}_{\coeffSHVec}$ for a smaller regularization parameter $\lambda$ than required by the least-squares method of reconstruction, with regularization and denoising LS-Regularized-Denoised. Thus, the proposed scheme better preserves features of the data, whereas the least-squares method of reconstruction over-smooths to achieve the same reconstruction accuracy. The reason for this is, as stated in Remark~\ref{Re:conditioning}, that the $\mathbf{P}_m$ matrices have been designed to be well-conditioned compared with the larger matrix $\mathbf{B}$ that is ill-conditioned with the number of samples equal to the number of coefficients, this results in a smaller regularization parameter required in Eq~\eqref{Eq:reg_nsht_ms} than Eq~\eqref{Eq:reg_LS_ms} for accurate reconstruction. 

The effects of this can be seen in the real data, where the ODFs in \figref{fig:odfs_proposed}(a) are sharper than \figref{fig:odfs_LS}(a) and the crossing fibers are more easily detected.  When the same $\lambda$ is used in the standard regularized least-squares reconstruction method as for the proposed method, the ODFs reconstructed using least-squares in \figref{fig:odfs_LS}(b) are noisy. This is particularly obvious in the voxels containing gray-matter which should be smooth. For example, in the four voxels enclosed by the yellow rectangle in \figref{fig:odfs_proposed}(b) and \figref{fig:odfs_LS}, for the ODFs obtained using the proposed method, in \figref{fig:odfs_proposed}(b), fibers can be clearly distinguished from peaks of the ODF. Whereas, for the ODFs obtained using least-squares, in \figref{fig:odfs_LS}(a), the over-smoothing of the ODFs means that some fibers cannot be distinguished and in \figref{fig:odfs_LS}(b) there are spurious peaks to the ODFs.

For multi-shell reconstruction, it can be seen in \figref{fig:NRMSE_spectral_error} that for all SNRs, fiber crossing angles and FAs the proposed multi-shell scheme (both with just regularization, nSPFt-Regularized, and regularization and denoising, nSPFt-Regularized-Denoised) has a smaller $\rm{NRMSE}_{\coeffSPFVec}$~(and also $\rm{NRMSE}_{\mathbf{d}}$) than the least-squares method of reconstruction (both with just regularization, LS-Regularized, and regularization and denoising, LS-Regularized-Denoised). For the least-squares method, $\rm{NRMSE}_{\coeffSPFVec}$ does not change much with SNR as the error is dominated by the ill-conditioned least-squares matrix inversion.

The proposed multi-shell scheme has a sampling structure which enables an accurate SPFt with the number of samples equal to the number of SH coefficients in each shell, unlike least-squares. The novel SPFt is also more flexible than least-squares in that it is able to have different SH band-limits for each shell. A benefit of least-squares is that it can be used with any sampling scheme, whereas the proposed scheme requires knowledge of the diffusion signal at specific points in $q$-space. However, the use of interpolation to approximate the value of the diffusion signal at these locations if measurements are taken on a different sampling grid could be explored as future work. 

For both single and multi-shell simulations in \figref{fig:NRMSE_ss_CA}, \figref{fig:NRMSE_ss_FA} and \figref{fig:NRMSE_spectral_error}, it is apparent that taking into account the non-Gaussian nature of the noise is more beneficial at low SNRs than at higher SNRs, with a bigger difference between the methods which solve the PML compared to just regularizing the solution for the both the proposed and least-squares methods. This is because the noise is approximately Gaussian at higher SNRs. Hence, why the ODFs generated from the real-data were indistinguishable when reconstruction was done using just regularization or when both regularization and denoising were used. The non-Gaussian noise modeling of the proposed scheme is particularly useful for low SNR data where Gaussian assumption is not valid, this occurs for example for high spatial resolution data. For moderate to high SNR data the regularized version of the proposed reconstruction algorithm is sufficient. 

 The proposed single and multi-shell sampling schemes allow accurate reconstruction of the diffusion signal with a reduced number of measurements and thus a shorter acquisition time. Hence, this new approach can be useful for the clinical diffusion MRI. The reconstructed diffusion signal can then be used with any of the many methods that compute features of the diffusion signal, such as computing the ODF.


\section*{Conclusions}\label{Sec:conclusions}
We have proposed novel single and multi-shell sampling and reconstruction schemes for diffusion MRI that have the minimum number of samples for reconstruction of the diffusion signal in the spherical harmonic (SH) and spherical polar Fourier (SPF) basis, respectively, with minimal assumptions while enabling accurate reconstruction of the diffusion signal. Robust and accurate reconstruction was achieved by regularizing the solution and accounting for the non-Gaussian nature of the noise by minimizing the penalized maximum likelihood~(PML) for Rician or non-central Chi noise. 

We evaluated the proposed single and multi-shell sampling schemes using synthetic and real data-sets by comparing their performance with that of the standard regularized least-squares method of reconstruction and regularized least-squares with PML to remove non-Gaussian noise. Both the single and multi-shell schemes have reconstruction algorithms that use smaller subsystems of linear equations to achieve better conditioning of the matrices used in the reconstruction algorithm compared to the standard least-squares method of reconstruction. In the single-shell case this enables accurate reconstruction with less regularization than the least-squares method of reconstruction. When the single-shell scheme was used on human brain data, this resulted in the orientation distribution functions (ODF) in the region of interest in the brain more clearly showing crossing fibers and having sharper peaks in single fiber areas compared to the least-squares method of reconstruction where the ODFs were over-smoothed.

In the case of the multi-shell sampling scheme the novel reconstruction algorithm, allowed by the separability of the SPF basis, enables different SH band-limits per shell, depending on the $b$-value of the shell, enabling accurate reconstruction with the minimum number of samples for reconstruction in the SPF basis to be achieved. This is not possible with least-squares, where the least-squares matrix becomes very poorly conditioned with this number of samples, resulting in a much higher reconstruction error than the proposed method. 

Considering the non-Gaussian nature of the noise also improves reconstruction accuracy at low SNR and is expected to be useful for reconstructing diffusion MRI data where this occurs, such as for high spatial resolution data. The proposed single and multi-shell sampling schemes are expected to be useful in clinical diffusion MRI where acquisition times need to be as short as possible.

\section*{Supporting information}


\paragraph*{S1 File.}
\label{S1_File}
{\bf Supplementary results.}  This file contains additional results to those presented in the Validation - Synthetic data subsection.

\section*{Acknowledgments}
We thank Dr. Divya Varadarajan and Dr. Justin P. Haldar for providing their code to their majorize-minimize framework presented in \cite{Varadarajan:2015}.

\nolinenumbers

\end{document}